\documentclass[prd,onecolumn,notitlepage,amsmath,amssymb,floatfix,superscriptaddress]{revtex4-1}
\usepackage[dvipdfmx]{graphicx}
\usepackage{dcolumn}
\usepackage{bm}
\usepackage{color}
\usepackage{hyperref}

\hypersetup{
  bookmarksnumbered=true,
  linkcolor=red,
  citecolor=green,
  urlcolor=cyan
}

\newcommand{\mnras}{Mon.~Not.~Roy.~Astron.~Soc.}
\newcommand{\physrep}{Phys.~Rep.}
\newcommand{\jcap}{JCAP}
\newcommand{\pasj}{PASJ}

\newcommand{\apjs}{\apj~Suppl.}
\newcommand{\aap}{Astronomy \& Astrophysics}

\newcommand{\bx}{{\bf x}}

\newcommand{\bs}{{\bf s}}
\newcommand{\bk}{{\bf k}}

\newcommand{\bq}{{\bf q}}

\newcommand{\bn}{{\bf n}}

\newcommand{\hk}{\hat{k}}
\newcommand{\hq}{\hat{q}}
\newcommand{\hn}{\hat{n}}

\newcommand{\tdelta}{\tilde{\delta}}
\newcommand{\tW}{\tilde{W}}

\newcommand{\br}{{\rm b}}
\newcommand{\dml}{\delta_{{\rm m}L}}
\newcommand{\tdml}{\tilde{\delta}_{{\rm m}L}}

\def\avrg#1{\left\langle #1 \right\rangle}

\begin{document}

\title[]{The impact of large-scale tides on cosmological distortions \\
via redshift-space power spectrum}

\author{Kazuyuki Akitsu}
\affiliation{Kavli Institute for the Physics and Mathematics of the Universe
(WPI), The University of Tokyo Institutes for Advanced Study (UTIAS),
The University of Tokyo, Chiba 277-8583, Japan}
\affiliation{Department of Physics, Graduate School of Science, The University of
Tokyo, 7-3-1 Hongo, Bunkyo-ku, Tokyo 113-0033, Japan}
\author{Masahiro Takada}
\affiliation{Kavli Institute for the Physics and Mathematics of the Universe
(WPI), The University of Tokyo Institutes for Advanced Study (UTIAS),
The University of Tokyo, Chiba 277-8583, Japan}

\date{\today}

\begin{abstract}
Although the large-scale perturbations beyond a finite-volume survey region are not direct observables, these affect measurements of 
clustering 
statistics of small-scale (sub-survey) perturbations in the large-scale structure, compared with the ensemble average,  
via the mode-coupling effect.
In this paper we show that the large-scale tides induced by scalar perturbations cause apparent anisotropic distortions in the redshift-space power spectrum of galaxies in a 
way depending on an alignment between the tides, the wavevector of small-scale modes and the line-of-sight direction.
Using the perturbation theory of structure formation, 
we derive the {\em response} function of the redshift-space power spectrum to the large-scale tides.
We then investigate the impact of the large-scale tides on estimation of cosmological distances and the
redshift-space distortion parameter via the measured redshift-space power spectrum for a hypothetical large-volume 
survey, based on the Fisher matrix formalism.
To do this, we treat the large-scale tides as a signal, 
rather than an additional source of the statistical errors, and show that a degradation in the parameter is restored if we can employ the prior on the rms amplitude
expected for the standard cold dark matter (CDM) model. 
We also discuss whether the large-scale tides can be constrained at an accuracy better than the CDM prediction, 
if the effects 
up to a larger wavenumber in the nonlinear regime can be included.
\end{abstract}

{\let\newpage\relax\maketitle}

\section{Introduction}
A number of wide-area and deep galaxy surveys are ongoing and planned, aimed at revealing the nature of primordial perturbations, 
the physics in the early universe, 
the curvature of the universe, the origin of the cosmic acceleration as well as weighing the neutrino mass via a measurement of large-scale structure probes such as weak gravitational lensing, 
baryon acoustic oscillations (BAO), 
galaxy clustering and redshift-space distortions \citep[e.g.,][]{Takadaetal:06,Takadaetal:14,TakadaDore:15}.
In particular, when combined with the high-precision measurement of cosmic microwave background (CMB) anisotropies,  large-scale structure probes allow one to study the time evolution of the perturbations over cosmic ages, which is sensitive to the aforementioned physics and cosmological parameters \citep[e.g.][]{Alametal:17}. 

The linear perturbation theory can accurately describe the time evolution of large-scale perturbations in structure formation, based on the standard
$\Lambda$ and cold dark matter dominated cosmology with Gaussian adiabatic initial conditions (hereafter $\Lambda$CDM) \cite{DodelsonBook}, which successfully reproduces 
the high-precision measurements of CMB anisotropies, yielding stringent constraints on the cosmological parameters \citep{PlanckCosmology:16}.
The linear theory, however, breaks down in the late-time universe, which is relevant for galaxy surveys, because the {\em nonlinear} structure formation 
induces a mode coupling between different Fourier modes of the perturbations owing to the nature of nonlinear, long-range gravity \citep[][for a thorough review]{Bernardeauetal:02,Desjacquesetal:16}. As a result, the power spectrum of large-scale structure probes, measured from a galaxy survey, no longer carries the full information unlike the CMB anisotropies, and the statistical properties display a substantial non-Gaussianity that is described by higher-order correlation functions \citep{TakadaJain:03a}. A better understanding of the nonlinear structure formation is thus 
required
in order to attain the full potential of wide-area galaxy surveys. 

Even though wide-area galaxy surveys are to cover a huge cosmological volume, there is an unavoidable uncertainty in the statistical analysis of large-scale structure probes arising due to finiteness of the survey volume as well as the nonlinear mode coupling, as studied in Refs.~\cite{TakadaHu:13,Hamiltonetal:06,Sefusattietal:06,TakadaBridle:07,TakadaJain:09,Satoetal:09,Baldaufetal:11,SherwinZaldarriaga:12,dePutteretal:12,TakadaSpergel:13,Schaanetal:14,Lietal:14a,Lietal:14b,Takahashietal:14,Manzottietal:14,CarronSzapudi:15,Daietal:15,Shirasakietal:17,Lietal:16,IpSchmidt:17,Akitsuetal:17,BarreiraSchmidt:17a}.
A finite-volume survey realization is generally embedded into large-scale perturbations that are not directly observable -- which we hereafter call ``super-survey modes'' \citep{TakadaHu:13,CarronSzapudi:15}. Although the super-survey modes have small amplitudes and are well in the linear regime for a wide-area galaxy survey, it causes a non-negligible effect on the small-scale perturbations due to the nonlinear mode coupling, compared with the statistical accuracies in measurements of the small-scale perturbations. Hence, it is necessary to include the effects in the cosmological analysis \citep[e.g.][]{Kuraseetal:17} in order not to have an biased estimation in cosmological parameters as well as not to have too optimistic cosmological constraints. 

The physical effects of super-survey modes on structure formation at equal time arise from the second-derivative tensor of the large-scale gravitational field 
due to the equivalence principle \citep{Baldaufetal:11,Akitsuetal:17}. The tensor is decomposed into two modes: the trace part or 
the large-scale density contrast and the trace-less tensor which we hereafter call
the large-scale tides. While the effect of the large-scale density contrast is well studied in previous studies \citep[e.g.][]{TakadaHu:13}, 
the effect of the large-scale tides has not been fully studied, except for some studies \citep{Schmidtetal:14,Daietal:15,IpSchmidt:17,Akitsuetal:17,BarreiraSchmidt:17a}. In our previous 
study \cite{Akitsuetal:17}, we showed that the large-scale tides cause an anisotropic clustering in the redshift-space power spectrum. 
The effect mimics the redshift-space distortion effect \cite{Kaiser:87,Hamilton:98} arising from the peculiar velocities of large-scale structure tracers as well as the 
Alcock-Paczynski (AP) distortion \cite{AlcockPaczynski:79,SeoEisenstein:03,HuHaiman:03} arising from the use of an incorrect cosmological model in the clustering analysis. Yet we studied only the partial effect,  the effect of the large-scale tides on the real-space clustering; in other words, we did not include the effect of the large-scale tides on the redshift-space distortion as well as a modulation in the mapping between real- and redshift-space distributions of galaxies.

Hence, the purpose of this paper is to study the effects of the large-scale tides on the redshift-space power spectrum, based on the perturbation theory \citep{Bernardeauetal:02}. To do this, we derive the {\em response} functions of the redshift-space power spectrum to
super-survey modes, which describe how the super-survey modes for a given survey realization affect the redshift-space power spectrum as a function of wavevector $\bk$ and the line-of-sight direction, say $\hat{\bn}$, relative to the large-scale tides.
We then discuss the impact of the tides on estimation of cosmological distances and the redshift-space distortion parameter via a measurement of the redshift-space power spectrum for a hypothetical large-volume galaxy survey, using the Fisher matrix formalism.

The rest of this paper is organized as follows. In \S\ref{sec:preliminaries}, we define the super-survey modes and introduce its 
isotropic component and anisotropic components.
In \S\ref{sec:response}, we derive the response function of the redshift-space power spectrum to super-survey modes by considering the squeezed-limit bispectrum that is a cross-correlation of the super-survey modes with the redshift-space power spectrum estimator. 
In \S\ref{sec:APtest}, we study the impact of the super-survey modes on cosmological parameter estimation from a measurement of the redshift-space power spectrum, including the AP test, 
based on the Fisher information matrix analysis.
\S\ref{sec:discussion} is devoted to the discussion.
In Appendix~\ref{app:multipole}, we give expressions for the multipole expansion of the 
redshift-space power spectrum in the presence of the super-survey modes.

\section{Preliminaries}
\label{sec:preliminaries}

The redshift-space density field of galaxies observed in a finite-volume survey region can be expressed, using 
the survey window function $W(\bx)$, following the formulation in Ref.~\cite{TakadaHu:13}:
\begin{equation}
\delta_{sW}(\bx)=W(\bx)\delta_{s}(\bx)
\end{equation}
where $\delta_{sW}(\bx)$ is the observed density field, $\delta_s(\bx)$ is the underlying true density field in redshift space, and 
the survey window is defined in that 
$W(\bx)=1$ if $\bx$ is inside the survey geometry, and otherwise $W(\bx)=0$. 
Throughout this paper 
we assume that a survey window is given in the background comoving coordinate.  
The survey volume is defined in terms of the survey window as
\begin{equation}
V_W=\int\!\!\mathrm{d}^3\bx~ W(\bx)
\end{equation}
In the following we assume a well-behaved survey window for simplicity; we do not consider effects of masks 
that might cause
additional mode-coupling between high-$k$ modes in the observed power spectrum. 
The Fourier transform of the density field is 
\begin{equation}
\tdelta_{sW}(\bk)=\int\!\!\frac{\mathrm{d}^3\bq}{(2\pi)^3}\tW(\bq)\tdelta_s(\bk-\bq),
\end{equation}
where quantities with tilde symble such as $\tdelta_s(\bk)$ are their Fourier transforms. 
The survey window $\tW(\bk)$ is non-vanishing for $k\ll 1/L$, while $\tW(\bk)\simeq 0$ for $k\gg 1/L$, 
where $L$ is a typical scale of the survey volume. 
The above equation explicitly shows that the Fourier transform of the observed field 
has a contribution from long-wavelength modes 
beyond a survey window, i.e. super-survey modes, via 
 a convolution with the survey window.

Redshift-space distortion (RSD) effect due to peculiar velocities of galaxies causes a modulation in the observed clustering pattern 
of galaxies along the line-of-sight direction. Thus the RSD effect violates a statistical isotropy of the galaxy distribution. 
For this reason the monopole power spectrum, measured by
the azimuthal-angle average of power spectrum over $\mathrm{d}\Omega_{\bk}$, 
cannot carry
the full information. Instead a standard approach to quantify the redshift-space clustering of galaxies
is using the power 
spectrum, given as a function of the three-dimensional wavevector $\bk$.
An estimator of the power spectrum for a given survey window is defined as
\begin{equation}
\hat{P}_s(\bk)\equiv \frac{1}{V_W}\int_{\bk'\in \bk}\frac{\mathrm{d}^3\bk'}{V_{\bk}}\tdelta_{sW}(\bk)\tdelta_{sW}(-\bk),
\label{eq:est_ps}
\end{equation}
where the integration is done over a volume element around the mode $\bk$ (a target wavevector 
for the power spectrum measurement), and $V_{\bk}$ is the volume: 
$V_{\bk}\equiv \int_{\bk'\in \bk}\!\!\mathrm{d}^3\bk'$. If a bin width around the bin $\bk$ is given by $\Delta k$, 
 $V_{\bk}\simeq (\Delta k)^3$. 
This definition does {\it not} include an angle average of $\mathrm{d}\Omega_{\bk}$, unlike a definition of the monopole
power spectrum. Hence, at this point, the redshift-space power spectrum $\hat{P}_s(\bk)$ is given as a function of the three-dimensional 
vector, $\bk$. A standard method usually further assumes the statistical isotropy in the two-dimensional plane (angular direction) perpendicular 
to the line-of-sight direction, and then uses the power spectrum given as a function of {\it two-dimensional} vector $(k_\parallel,k_\perp)$,  where 
$k_{\parallel}$ is the line-of-sight direction component of $\bk$,  $\bk_\perp$ is the vector in the two-dimensional plane perpendicular to the line-of-sight direction, and $k_\perp=|\bk_\perp|$.
Here we do not introduce the angle average over $\mathrm{d}\varphi_{\bk_\perp}$ in the perpendicular plane, 
where $\varphi_{\bk_\perp}$ is defined via 
$\bk_{\perp}=k_\perp\left(\cos\varphi_\bk,\sin\varphi_\bk\right)$, and keep the general definition of $\hat{P}_s(\bk)$ because 
the large-scale tides generally cause anisotropic distortions in the redshift-space clustering pattern of galaxies in all three-dimensional directions \citep[also see Ref.][for the similar discussion]{Shiraishietal:17}. 

Given the definition of the redshift-space power spectrum,
\begin{equation}
\avrg{\tdelta_s(\bk)\tdelta_s(\bk')}\equiv (2\pi)^3P_s(\bk)\delta^{3}_D(\bk+\bk'),
\end{equation}
where $\delta_D^3(\bk)$ is the Dirac delta function, the ensemble average of the estimator (Eq.~\ref{eq:est_ps}) is found to be an unbiased estimator
of the underlying power spectrum for modes with $k\gg 1/L$:
\begin{eqnarray}
\avrg{\hat{P}_s(\bk)}&=&\frac{1}{V_W}\int_{\bk\in \bk'}\!\frac{\mathrm{d}^3\bk'}{V_{\bk}}
\int\!\!\frac{\mathrm{d}^3\bq}{(2\pi)^3}\left|\tW(\bq)\right|^2P_s(\bk-\bq)\nonumber\\
&\simeq& \frac{1}{V_W}\int_{\bk\in \bk'}\!\frac{\mathrm{d}^3\bk'}{V_{\bk}}P_s(\bk)
\int\!\!\frac{\mathrm{d}^3\bq}{(2\pi)^3}\left|\tW(\bq)\right|^2\nonumber\\
&\simeq & \frac{P_s(\bk)}{V_W}\int_{\bk\in \bk'}\!\frac{\mathrm{d}^3\bk'}{V_{\bk}}
\int\!\!\frac{\mathrm{d}^3\bq}{(2\pi)^3}\left|\tW(\bq)\right|^2=P_s(\bk).
\end{eqnarray}
Here we used $P_s(\bk-\bq)\simeq P_s(\bk)$ over the integration rage of $\mathrm{d}^3\bq$ which the window function supports and assumed that $P_s(\bk)$
is not a rapidly  varying function within the $\bk$-bin. In addition, we used the general identity 
for the window function \cite{TakadaHu:13}:
\begin{equation}
V_W=\int\!\!\mathrm{d}^3\bx~W(\bx)^n=\int\!\!\left[\prod_{a=1}^{n}\frac{\mathrm{d}^3\bq_a}{(2\pi)^3}\tW(\bq_a)\right](2\pi)^3\delta_D^{3}(\bq_{1\dots n}),
\end{equation}
where $\bq_{1\dots n}\equiv \bq_1+\bq_2+\cdots+\bq_n$.

Similarly to Takada \& Hu \cite{TakadaHu:13} and Akitsu et al. \cite{Akitsuetal:17}, 
we study effects of super-survey modes on the redshift-space power spectrum. The super-survey modes we focus on are
 the large-scale density contrast and the large-scale tides, defined 
in terms of the {\it linear} matter density fluctuation field
as
\begin{eqnarray}
\delta_\br&\equiv& \frac{1}{V_W}\int\!\!\mathrm{d}^3\bx~ W(\bx)
\tdml(\bx)
=\frac{1}{V_W}\int\!\!\frac{\mathrm{d}^3\bq}{(2\pi)^3}
\tdml(\bq)\tW(-\bq), \nonumber\\
\tau_{ij} &\equiv & \frac{1}{4\pi G\bar{\rho}_{\rm m} a^2 V_W}\int\!\!\mathrm{d}^3\bx~W(\bx)\left[
\Phi_{,ij}(\bx)-\frac{\delta^K_{ij}}{3}\nabla^2\Phi(\bx)\right]
=\frac{1}{V_W}\int\!\!\frac{\mathrm{d}^3\bq}{(2\pi)^3}\left(\hat{q}_i\hat{q}_j-\frac{\delta^K_{ij}}{3}\right)\tdml(\bq)\tW(-\bq), 
\end{eqnarray}
where $\hq_i\equiv q_i/q$, $\hq_i \hq^i=1$, $\delta^K_{ij}$ is the Kronecker delta, and $\Phi(\bx)$ is the 
gravitational potential field. Here 
we assumed that a survey volume is sufficiently large, and therefore the matter density field contributing
 the super-survey modes are 
in the linear regime, denoted as $\dml(\bx)$. Under this setting, $|\delta_\br|, |\tau_{ij}|\ll 1$.
These super-survey modes are not direct observables and vary with survey realizations. For a particular 
survey realization, $\delta_\br, \tau_{ij}$ have 
particular constant values.
The {\em expectation} values of the ensemble averages, i.e. the averages over different, possible survey realizations for a fixed volume, 
are computed if the linear matter power spectrum at long wavelengths for super-survey modes,  
$P^L(k)$, 
is given for a given cosmological model:
$\avrg{\delta_\br}=\avrg{\tau_{ij}}=0$, and the variances are 
\begin{eqnarray}
&&\sigma_b^2\equiv \langle \delta_\br^2 \rangle = \frac{1}{V_W^2}\int \frac{d^3 \bq}{(2\pi)^3} 
P^L(q) \left| W(\bq) \right|^2, \nonumber \\ 
&&\langle \delta_\br \tau_{ij} \rangle = \frac{1}{V_W^2}\int \frac{d^3 \bq}{(2\pi)^3} 
\left( \hat{q}_i \hat{q}_j -\frac{1}{3}\delta^{K}_{ij} \right)P^L(q) \left| W(\bq) \right|^2,   \nonumber\\
&&\langle \tau_{ij} \tau_{lm} \rangle =  \frac{1}{V_W^2}\int \frac{d^3 \bq}{(2\pi)^3} 
\left( \hat{q}_i \hat{q}_j -\frac{1}{3}\delta^{K}_{ij} \right)\left( \hat{q}_l \hat{q}_m -\frac{1}{3}\delta^{K}_{lm} \right)
P^L(q) \left| W(\bq) \right|^2, \label{eq:sigma_tau} 
\end{eqnarray}
where $P^L(q)$ is the linear matter power spectrum. 

In this paper we consider an isotropic window for simplicity; $\tW(\bq)=\tW(q)$. In this case
the variances of large-scale tides are simplified as
\begin{eqnarray}
&&\avrg{\delta_\br\tau_{ij}}=0,\nonumber\\
&&\sigma_\tau^2\equiv \avrg{(\tau_{11})^2}=\avrg{(\tau_{22})^2}=\avrg{(\tau_{33})^2}=\frac{3}{4}\avrg{(\tau_{ij})^2}_{i\ne j}
=\frac{4}{45V_W^2}\int\!\!\frac{q^2\mathrm{d}q}{2\pi^2}P^L(q)\left|\tW(q)\right|^2=\frac{4}{45}\sigma_\br^2.
\end{eqnarray}

\section{Responses of redshift-space power spectrum to super-survey modes}
\label{sec:response}

\subsection{Redshift-space distortion effects}

In a redshift galaxy survey, 
the radial position of each galaxy 
needs to be inferred from its observed redshift. Here 
the observed redshift can be  modified by a peculiar velocity of the galaxy through the Doppler effect, 
causing an apparent displacement of the inferred galaxy position from the true position:
\begin{equation}
\bs=\bx + \frac{v_{\parallel}(\bx)}{\mathcal{H}(z)}\hat{\bn}, 
\label{eq:s_x}
\end{equation}
where $\bs$ is the inferred position of a galaxy in redshift space, $\bx$
is the true position in real space, $v_{\parallel}$ is the radial component of the peculiar velocity,
$\mathcal{H}(z)$ is the comoving Hubble rate, and $\hat{\bn}$ is the unit vector of 
the line-of-sight direction. 
With this coordinate transformation, the density field in redshift space can be expressed as
\begin{equation}
\rho_s(\bs) = \int\!\!\mathrm{d}^3\bx~\rho(\bx)~\delta_D^{3}\! \left(\bs - \bx -\frac{v_{\parallel}(\bx)}{\mathcal{H}(z)}\hat{\bn}\right) , 
\label{eq:rho_s-r}
\end{equation}
where $\rho_s(\bs)$ or $\rho(\bx)$ denotes the redshift- or real-space density field of galaxies, respectively.
In the following quantities with subscript ``$s$'' denote their redshift-space 
quantities.
Fourier-transforming Eq.~(\ref{eq:rho_s-r}), $\int\!\mathrm{d}^3\bs~e^{i\bk\cdot\bs}$, yields
%
\begin{equation}
\delta_D^{3}(\bk) + \tdelta_s(\bk) = \int\!\!\mathrm{d}^3 \bx ~ [1+\delta(\bx)] e^{-i\bk\cdot\bx - i(\bk \cdot \hat{\bn})\frac{v_{\parallel}}{\mathcal{H}}} .
\label{eq:delta_s-r_exact}
\end{equation}
This transformation is exact even if multiple galaxies are mapped to the same position in redshift space, which can happen, e.g. in a nonlinear high-density region.
Such multi-streaming regions are 
beyond the scope of this paper, 
and we ignore the effects in this paper for simplicity. 
In this setting we can rewrite Eq.~(\ref{eq:delta_s-r_exact}) as 
\begin{eqnarray}
\tdelta_s(\bk) &=& \int\!\!\mathrm{d}^3 \bx \left[1+\delta(\bx)-\left|\frac{\partial s_i}{\partial x_j}\right| \right]
 e^{-i\bk\cdot\bx - i(\bk \cdot \hat{\bn})\frac{v_{\parallel}}{\mathcal{H}}} \nonumber \\
&\simeq & \int\!\!\mathrm{d}^3 \bx \left[\delta(\bx) - \frac{1}{\mathcal{H}(z)}\frac{\partial v_{\parallel}}{\partial \hat{\bn}}\cdot\hat{\bn}\right]
  e^{-i\bk\cdot\bx - i(\bk \cdot \hat{\bn})\frac{v_{\parallel}}{\mathcal{H}}},
 \label{eq:delta_s-r}
\end{eqnarray}%
where we kept the peculiar velocity up to the linear order in an expansion of the Jacobian, $|\partial s_i/\partial x_j|$, assuming 
$|v_\parallel/{\cal H}|\ll 1$.

Using the perturbation theory of structure formation \cite{Bernardeauetal:02}, we can 
express the redshift-space density field $\tdelta_s(\bk)$ in terms of the linear matter density field
$\tdml(\bk)$ as
\begin{equation}
\tdelta_s(\bk; t)\equiv \sum_{n=1}^\infty 
\int\left[\prod_{a=1}^{\infty}
\frac{\mathrm{d}^3\bk_a}{(2\pi)^3}\right]Z_i(\bk_1,\dots,\bk_i)
\tdml(\bk_1,t)\cdots\tdml(\bk_i,t)(2\pi)^3\delta_D^{3}(\bk_{1\dots i}-\bk)
\end{equation}
where we have introduced the notation, $\bk_{12\dots i}\equiv \bk_1+\bk_2+\cdots +\bk_i$, 
and $Z_i(\bk_1,\dots,\bk_i)$ is the mode-coupling 
kernel between different Fourier modes with $\bk_1, \dots ,\bk_i$. 
 We throughout this paper 
employ a distant observer approximation for simplicity. 
In the following discussion we use the density fields
up to the second-order, which are given as 
\begin{eqnarray}
\tdelta_s(\bk)&\simeq& \tdelta_s^{(1)}(\bk)+\tdelta_s^{(2)}(\bk)\nonumber\\
&=&Z_1(\bk)\tdml(\bk)+\int\!\!\frac{\mathrm{d}^3\bk_1}{(2\pi)^3}\frac{\mathrm{d}^3\bk_2}{(2\pi)^3}
Z_2(\bk_1,\bk_2)\tdml(\bk_1,t)\tdml(\bk_2,t)(2\pi)^3\delta^{3}_D(\bk_{12}-\bk).
\label{eq:deltas_pt}
\end{eqnarray}
Using the standard Eulerian perturbation theory \cite{Goroffetal:86,Makinoetal:92,JainBertschinger:94}, where 
an irrotational, pressure-less single-fluid matter field is assumed, the kernels are given as
\begin{eqnarray}
&&Z_1(\bk)\equiv b+f\mu^2, \label{eq:delta_kaiser}\nonumber\\
&&Z_2(\bk_1,\bk_2)
\equiv bF_2(\bk_1,\bk_2)+f\mu^2 G_2(\bk_1,\bk_2)+\frac{f\mu k}{2}\left[
\frac{\mu_1}{k_1}\left(b+f\mu_2^2\right)+
\frac{\mu_2}{k_2}\left(b+f\mu_1^2\right)
\right],
\label{eq:Zn}
\end{eqnarray}
where $\bk\equiv \bk_1+\bk_2$,
$\mu$ is the cosine angle between the wavevector $\bk$ and the line-of-sight direction, 
$\mu\equiv \hat{\bn}\cdot\hat{\bk}=k_\parallel/k$ ($k_{\parallel}$ is the component along the line-of-sight direction),
$f\equiv \mathrm{d}\ln D/\mathrm{d}\ln a$,  $D$ is the linear growth rate, and $b$ is the linear bias parameter of galaxies. 
The pioneer work for the RSD effect is given in Ref.~\cite{Kaiser:87},
and see Refs.~\cite{Hivonetal:95,Verdeetal:98,Scoccimarroetal:99,Scoccimarro:04} for the extension to the higher-order terms.   
Throughout this paper we assume the linear galaxy bias to model how the real-space distribution of galaxies is related to that
of matter. Although the effect of the large-scale tides could cause an additional biasing effect on the tracers \cite{McDonaldRoy:09,Desjacquesetal:16,Chanetal:12,Saitoetal:14}, 
the effect on the power spectrum is of the order of $\mathcal{O}\left((\dml)^2\right)$, compared with the $\mathcal{O}(\dml)$ effect in $b$, so we ignore the effect for simplicity.  
The kernels $F_2(\bk_1,\bk_2)$ and $G_2(\bk_1,\bk_2)$ are the second-order 
kernels for the density and velocity perturbations, given by Eqs.~(45) and (46) in Ref.~\cite{Bernardeauetal:02}:
\begin{eqnarray}
&&F_2(\bk_1,\bk_2)=\frac{5}{7}+\frac{1}{2}\left(\frac{1}{k_1^2}+\frac{1}{k_2^2}\right)\left(\bk_1\cdot\bk_2\right)
+\frac{2}{7}\frac{\left(\bk_1\cdot\bk_2\right)^2}{k_1^2k_2^2},\nonumber\\
&&G_2(\bk_1,\bk_2)=\frac{3}{7}+\frac{1}{2}\left(\frac{1}{k_1^2}+\frac{1}{k_2^2}\right)\left(\bk_1\cdot\bk_2\right)
+\frac{4}{7}\frac{\left(\bk_1\cdot\bk_2\right)^2}{k_1^2k_2^2}.
\end{eqnarray}

\subsection{Derivation of the responses of redshift-space power spectrum to super-survey modes}

We now consider how super-survey modes affect the redshift-space power spectrum observed in a finite-volume 
survey. Following the discussion in Refs.~\cite{TakadaHu:13,Akitsuetal:17}, in the presence 
of super-survey modes ($\delta_\br,\tau_{ij}$) for a given survey realization, the ``observed'' redshift-space 
power spectrum is formally expressed as
\begin{eqnarray}
{P}_{sW}(\bk; \delta_\br, \tau_{ij}) = {P}_{s}(\bk) + 
\frac{\partial {P_s(\bk)}}{\partial \delta_\br} \delta_\br+ \frac{\partial {P_s(\bk)}}{\partial \tau_{ij}}\tau_{ij}   
\label{eq:P_sW}.
\end{eqnarray}
Here we omitted the dependence of $P_{sW}(\bk)$ on the line-of-sight direction, $\hat{\bn}$, for notational simplicity and 
we explicitly denote that the observed spectrum $P_{sW}(\bk; \delta_\br,\tau_{ij})$ depends 
on the super-survey modes of a given survey realization, 
and $P_s(\bk)$ is the power spectrum without the super-survey modes. The functions $\partial  P_s(\bk)/\partial \delta_\br$
and $\partial P_s(\bk)/\partial \tau_{ij}$ are so-called 
``response'' functions describing a response of the redshift-space power spectrum 
to the super-survey modes via mode couplings in the nonlinear structure formation. 

Now we derive the response function using the perturbation theory.
The simplest way to do this is considering the squeezed limit of the bispectrum that arises from correlations 
between two short modes and one long mode (corresponding to the super-survey modes) \cite{Chiangetal:14}. 
More specifically, 
let us consider a correlation of $\hat{P}_{sW}(\bk)$ (Eq.~\ref{eq:est_ps}) with the large-scale matter density field, 
$\tdml(\bq)$ ($\bq$ is the long mode):
\begin{eqnarray}
\avrg{\hat{P}_{sW}(\bk)\tdml(\bq)}&=&\frac{1}{V_W}\int_{\bk'\in \bk}\!\!\frac{\mathrm{d}^3\bk'}{V_\bk}
\int\!\!\frac{\mathrm{d}^3\bq_1}{(2\pi)^3}\frac{\mathrm{d}^3\bq_2}{(2\pi)^3}
\avrg{\tdelta_s(\bk'-\bq_1)\tdelta_s(-\bk'-\bq_2)\tdml(\bq)}\tW(\bq_1)\tW(\bq_2)\nonumber\\
&=& \frac{1}{V_W}\int_{\bk'\in \bk}\!\!\frac{\mathrm{d}^3\bk'}{V_\bk}
\int\!\!\frac{\mathrm{d}^3\bq_1}{(2\pi)^3}\frac{\mathrm{d}^3\bq_2}{(2\pi)^3}
B_{ss{\rm m}}(\bk'-\bq_1,-\bk'-\bq_2,\bq)(2\pi)^3\delta_D^3(\bq_{12}-\bq)\tW(\bq_1)\tW(\bq_2),
\label{eq:ps_dq}
\end{eqnarray}
where we have defined the bispectrum between the redshift-space density field and the real-space density field:
\begin{equation}
\avrg{\tdelta_s(\bk_1)\tdelta_s(\bk_2)\tdml(\bq)}\equiv B_{ss{\rm m}}(\bk_1,\bk_2,\bq)(2\pi)^3\delta_D^3(\bk_1+\bk_2+\bq).
\end{equation}

For the case that $k\gg q_1,q_2,q$, 
the bispectrum in the above equation arises from so-called squeezed triangles where two sides are nearly equal and in opposite direction. 
To see this, we can make the variable changes $\bk-\bq_1\leftrightarrow \bk$ and $\bq_1+\bq_2\leftrightarrow \bq$ 
under the delta function condition $\bq_{12}+\bq={\bf 0}$ and the approximation that $k\ll q$. The bispectrum we are interested in reads
\begin{equation}
\lim_{q\rightarrow 0}B_{ss{\rm m}}(\bk,-\bk-\bq,\bq).
\end{equation}
In this limit, the triangle configuration describes how the redshift-space power spectrum $P_s(\bk)$ is modulated 
by the super-survey modes $\tdml(\bq)$. For convenience of the following discussion, 
with the help of Eq.~(\ref{eq:P_sW})
we assume  that the squeezed bispectrum can be 
described by the response of $P_s(\bk)$ to the super-survey modes as
\begin{equation}
\lim_{q\rightarrow 0}B_{ss{\rm m}}(\bk,-\bk-\bq,\bq)\equiv \left[\frac{\partial P_s(\bk)}{\partial \delta_b}+
\left(\hq_i\hq_j-\frac{\delta^K_{ij}}{3}\right)\frac{\partial P_s(\bk)}{\partial \tau_{ij}}\right] P^L(q).
\label{eq:B_squeezed}
\end{equation}
From Eq.~(\ref{eq:B_squeezed}), we can derive the response function 
$\partial P_s(\bk)/\partial \delta_\br$
from the 
angle average of the squeezed bispectrum over $\mathrm{d}^3\bq$ as

\begin{eqnarray}
&& \frac{\partial P_s(\bk)}{\partial \delta_b}P^L(q)\simeq
\lim_{q\rightarrow 0} \int\!\!\frac{\mathrm{d}\Omega_\bq}{4\pi}~ 
B_{ss{\rm m}}(\bk,-\bk-\bq,\bq),
\label{eq:response_angave}
\end{eqnarray}
With this derivation, the response to the large-scale tide, $\partial P_s(\bk)/\partial \tau_{ij}$ can be found from 
\begin{equation}
\frac{\partial P_s(\bk)}{\partial \tau_{ij}} \longleftarrow 
\mbox{coefficients in } \left(\hq_i\hq_j-\frac{\delta^K_{ij}}{3}\right)P^L(q)\mbox{ in }
\lim_{q\rightarrow 0}B_{ss{\rm m}}(\bk,-\bk-\bq,\bq).
\label{eq:response_tau}
\end{equation}

Using the perturbation theory ansatz for $\tdelta_s(\bk)$ (Eq.~\ref{eq:deltas_pt}) 
and assuming that the large-scale mode 
$\tdml(\bq)$ is in the linear regime, the leading-order contribution of 
the squeezed bispectrum can be expressed in terms
of the mode-coupling kernels as
\begin{eqnarray}
B_{ss{\rm m}}(\bk,-\bk-\bq,\bq)&\simeq& 2Z_1(\bk+\bq)Z_2(\bk+\bq,-\bq)
P^L(|\bk+\bq|)P^L(q)
+2Z_1(\bk)Z_2(\bk,\bq)P^L(k)P^L(q).
\label{eq:Bssm_expand}
\end{eqnarray}
Inserting Eq.~(\ref{eq:Zn}) into Eq.~(\ref{eq:Bssm_expand}) and using the relations (Eq.~\ref{eq:response_angave} and Eq.~\ref{eq:response_tau}), 
we can find 
that the response functions for the redshift-space power spectrum are
\begin{eqnarray}
\frac{\partial P_s(\bk)}{\partial \delta_\br}&=&
\left[\frac{47}{21}-\frac{1}{3}\frac{\mathrm{d}\ln P^L(k)}{\mathrm{d}\ln k}\right]b^2P^L(k)
+
\left[
\frac{b}{3}+\mu^2\left(\frac{26}{7}+2b\right)
-\frac{\mu^2}{3}(2+b)\frac{\mathrm{d}\ln P^L(k)}{\mathrm{d}\ln k}
\right]bf P^L(k)\nonumber\\
&&+
\left[
\frac{1}{21}\left(31+70b\right)-\frac{1}{3}(1+2b)\frac{\mathrm{d}\ln P^L(k)}{\mathrm{d}\ln k}
\right]f^2\mu^4 P^L(k)
+
\left[\frac{1}{3}(4\mu^2-1)-\frac{\mu^2}{3}\frac{\mathrm{d}\ln P^L(k)}{\mathrm{d}\ln k}\right]
f^3\mu^4P^L(k),
\label{eq:dpsddb}
\end{eqnarray}
and
\begin{eqnarray}
\frac{\partial P_s(\bk)}{\partial \tau_{ij}}&=&
\left[\frac{8}{7}\hk_i\hk_j -\hk_i\hk_j\frac{\mathrm{d}\ln P^L(k)}{\mathrm{d}\ln k}\right]b^2P^L(k)
\nonumber\\
&&
+\left[b\hat{n}_i\hat{n}_j+\frac{24}{7}\mu^2\hk_i\hk_j
-\mu\left(2\mu \hk_i\hk_j+ bh_{ij}\right)\frac{\mathrm{d}\ln P^L(k)}{\mathrm{d}\ln k}\right]bf P^L(k)
\nonumber\\
&&+
\left[
\frac{16}{7}\mu\hk_i\hk_j+4bh_{ij}
-\left(\mu\hk_i\hk_j+2bh_{ij}\right)\frac{\mathrm{d}\ln P^L(k)}{\mathrm{d}\ln k}
\right]f^2\mu^3P^L(k)
\nonumber\\
&&+
\left[
\left(4\mu h_{ij}-\hat{n}_i\hat{n}_{j}\right)
-\mu h_{ij}\frac{\mathrm{d}\ln P^L(k)}{\mathrm{d}\ln k}
\right]f^3\mu^4P^L(k),
\label{eq:dpsdtau}
\end{eqnarray}
where 
\begin{equation}
h_{ij}\equiv \hk_{(i} \hn_{j)}
= \frac{1}{2}
\left(\hk_i\hn_j+\hk_j\hn_i\right).
\end{equation}
These are full expressions of the responses of redshift-space power spectrum to the large-scale perturbations. Compared with 
the results in Ref.~\cite{Akitsuetal:17}, there are additional effects of the super-survey modes on 
the redshift-space power spectrum, that is,
there are terms including the couplings between the large-scale tides $\tau_{ij}$ and the line-of-sight direction $\hat{\bn}$ 
as expected.
The response function for $\delta_\br$, $\partial P_s(\bk)/\partial \delta_\br$, agrees with Eq.~(65)
in Ref.~\cite{NishimichiValageas:15} if we set $b=1$ in the above equation. 
The response functions, $\partial P_s(\bk)/\partial \delta_{\rm b}$ and $\partial P_s(\bk)/\partial \tau_{ij}$, 
show several effects caused by the super-survey modes.
First, the large-scale perturbations could speed up or slow down the growth of short modes:
for example, if the large-scale tide along a particular direction is positive, 
say $\tau_{ii}>0$, the expansion of a local volume along the direction is slower than that of the global universe, 
so the growth of short modes with $\bk$ along the direction can be enhanced. Secondly, the super-survey modes cause
a dilation of the comoving wavelengths. Because the large-scale perturbations can be realized as a modification of the local 
expansion, the comoving wavelengths which an observer infer are modulated by the super-survey modes, which
imprints a modulation in the power spectrum.
Thirdly, the super-survey modes alter
the peculiar velocities through the effects on the gravitational force, so alter the redshift-space distortion effects along 
the line-of-sight direction. 
Here the large-scale tides cause modifications in the clustering pattern along all the three directions. 

In the following we focus on the response function for $\tau_{ij}$, and we do not consider the response for $\delta_\br$.
From Eq.~(\ref{eq:dpsdtau}) we can find several types of anisotropies in the redshift-space power spectrum:
the standard RSD effect $\mu^2=\hk_i\hk_j\hn_i\hn_j$ (Kaiser factor), and the effects due to $\tau_{ij}$ that have dependences of
$\tau_{ij}\hk_i\hk_j$, $\tau_{ij}\hk_i\hn_j$, and $\tau_{ij}\hn_i\hn_j$, respectively.
First, let us remind of the physical origin of the Kaiser factor.
It
comes from $\partial_i v_j \hn_i \hn_j$ (see Eq.~\ref{eq:delta_s-r}).
This means that the Kaiser anisotropy reflects the projection of the velocity shear ($\partial_{(i} v_{j)}$, in Fourier space $\propto \hk_i\hk_j$)
onto the line-of-sight direction.
In other words, since the velocity shear corresponds to the tidal field,
the Kaiser factor can be interpreted as the projection of the short-mode tidal field onto the line-of-sight direction.
Next, the terms proportional to $\tau_{ij}\hk_i\hk_j$ represent a coupling between the
large-scale tides $\tau_{ij}$ and the small-scale tides, where the latter has directional dependences given by
 $\propto \left( \hk_i \hk_j -\frac{1}{3}\delta_{ij}^K \right)$. The terms of
$\tau_{ij} \hn_i\hn_j$ are like the Kaiser factor, that is, the projection of the large-scale tides $\tau_{ij}$
onto the line-of-sight direction.
Note that the terms proportional to $h_{ij}$ always appear with $\mu = \hat{\bk} \cdot \hat{\bn}$,
because of the parity invariance of the power spectrum, i.e. $P_s(\bk)=P_s(-\bk)$.
Then, $\tau_{ij} \hk_i \hn_j \mu = \tau_{ij} \hk_i \hk_l \hn_j \hn_l$ is a consequence of the projection of the coupling between 
the large-scale tides $\tau_{ij}$ and the  small-scale velocity $\propto \hk_i$ onto the line-of-sight direction.

\subsection{\texorpdfstring{The large-scale mode effects on the two-dimensional redshift-space power spectrum:
$P^{\rm 2D}_s(k_\parallel,k_\perp)$}{The large-scale mode effects on the two-dimensional redshift-space power spectrum}  }

The main purpose of this paper is to estimate the impact of super-survey modes on the RSD measurements as well as
Alcock-Paczynski (AP) test \cite{AlcockPaczynski:79} through a measurements of the redshift-space power spectrum. 
To do this, we employ the standard approach used in an analysis of the redshift-space power spectrum. Since the RSD
effect is only along the line-of-sight direction and does not affect the clustering pattern in the two-dimensional 
plane perpendicular to the line-of-sight direction, a usual way to measure the redshift-space power spectrum is making 
the angle average given as
\begin{equation}
{P}^{\rm 2D}_{sW}(k_\parallel,k_\perp; \tau_{ij})\equiv \int^{2\pi}_0\!\!\frac{\mathrm{d}\varphi_{\bk_\perp}}{2\pi}~{P}_{sW}(\bk;
\tau_{ij}),
\label{eq:p2d_def}
\end{equation}
where we have set the line-of-sight direction as $z$-axis, $\hn_i = \delta^K_{iz}$ and used the decomposition of wavevector, $\bk=(k_\perp \cos\varphi_{\bk_\perp},k_\perp\sin\varphi_{\bk_\perp},k_\parallel)$
with the conditions $(k_\perp,k_\parallel)=k\left(\sqrt{1-\mu^2},\mu\right)$. 

By inserting Eqs.(\ref{eq:P_sW}) and (\ref{eq:dpsdtau}) into Eq.~(\ref{eq:p2d_def}) we can find
\begin{eqnarray}
P^{\rm 2D}_{sW}(k_\perp,k_\parallel; \tau_{ij})&=& (b+f\mu^2)^2P^L(k)+
\left[\frac{8}{7}b^2 P^L(k)-b^2\frac{\mathrm{d}P^L(k)}{\mathrm{d}\ln k}\right]\frac{3\mu^2-1}{2}\tau_{33}
\nonumber\\
&&+fb\left[\left\{b+ \frac{12}{7}\mu^2 (3\mu^2-1)\right\}P^L(k)
-\mu^2\left\{b+(3\mu^2-1)\right\}\frac{\mathrm{d}P^L(k)}{\mathrm{d}\ln k}\right]\tau_{33}
\nonumber\\
&&+f^2\mu^4\left[
\left\{4b+\frac{8}{7}(3\mu^2-1)\right\}P^L(k)
-\left(2b+\frac{3\mu^2-1}{2}\right)\frac{\mathrm{d}P^L(k)}{\mathrm{d}\ln k}\right] \tau_{33}
\nonumber\\
&&+f^3\mu^4\left[
\left(4\mu^2-1\right)P^L(k)
-\mu^2\frac{\mathrm{d}P^L(k)}{\mathrm{d}\ln k}
\right]\tau_{33},
\label{eq:p2d_sW}
\end{eqnarray}
where we have used the following identities under the presence of the line-of-sight direction
\begin{eqnarray}
&&\int_0^{2\pi}\!\!\frac{\mathrm{d}\phi_{\bk_\perp}}{2\pi} \hk_i\hk_j  =
\ \frac{1-\mu^2}{2}\delta_{ij}^K
+\frac{3\mu^2-1}{2}\hat{n}_i\hat{n}_j, \nonumber \\
&&\int_0^{2\pi}\!\!\frac{\mathrm{d}\phi_{\bk_\perp}}{2\pi} \hk_i
=\mu \hat{n}_i,
\end{eqnarray}
with the trace-less condition of $\tau_{ij}$, i.e. $\tau_{ij}\delta_{ij}^K=0$.
Eq.~(\ref{eq:p2d_sW}) is one of the main results of this paper. The equation shows that the large-scale tides
cause an additional anisotropic clustering in the two-dimensional redshift-space power spectrum in addition to the 
Kaiser distortion. 
The amount of the 
distortion depends on the 
line-of-sight component of the tides, $\tau_{33}$, in a given survey realization. The tides cause 
anisotropic distortions up to the order of $\mu^6$, while the standard Kaiser RSD effect causes distortions up to 
$\mu^4$. 
Thus the large-scale tides in a given survey realization 
cause a bias in the redshift-space power spectrum. There are two ways to take into account 
the effect. One way is to include the effect as an additional noise in the error covariance matrix of the 
power spectrum as studied in Ref.~\cite{Akitsuetal:17}. Alternative approach, which we take in this paper, 
is to treat the effect as a signal rather than noise. We can model this effect by treating the bias 
as a purely systematic additive shift in the redshift-space power spectrum, where an amount of the bias is 
given by the power spectrum response multiplied by a free parameter $\tau_{33}$. Then we can use 
the measured power spectrum to infer the $\tau_{33}$ value in the survey realization. We will study how a large-volume galaxy redshift survey can constrain the large-scale tides and also how it could cause a degradation in cosmological parameters.

\section{The impact of large-scale tidal effect on redshift-space power spectrum}
\label{sec:APtest}

\subsection{Fisher information matrix}
\label{sec:Fisher}

In this section, following Refs.~\cite{SeoEisenstein:03} and \cite{HuHaiman:03}, 
we study how the large-scale tides affect the BAO and RSD measurements 
in the redshift-space power spectrum \citep{Takadaetal:14}, based on the Fisher information matrix formalism. 

The two-point correlation function of galaxies is measured
as a function of the separation lengths between paired galaxies.
To measure this separation, the position of each galaxy needs to be inferred 
from the measured redshift and angular position.  
Then the separation lengths perpendicular and
parallel to the line-of-sight direction from the measured quantities are
given as $r_\perp\propto \Delta \theta$ and $r_\parallel\propto \Delta
z$, with $\Delta \theta $ and $ \Delta z$ being the differences between
the angular positions and the redshifts of the paired galaxies.
To convert the observables ($\Delta \theta$, $\Delta z$) to the quantities
$(r_\perp,r_\parallel)$, one has to assume a reference cosmological model.
Considering this transformation, the wavenumbers are given as
\begin{equation}
k_{\perp,{\rm ref}}= \frac{D_{A}(z)}{D_{A, {\rm
 ref}}(z)}
k_{\perp},
\hspace{1em}
k_{\parallel, {\rm ref}}=\frac{H_{\rm ref}(z)}{H(z)}
k_\parallel,
\end{equation}
where $D_A(z)$ is the angular diameter distance and $H(z)$ is the Hubble expansion rate.
The quantities with subscript ``ref'' mean the quantities for 
 an assumed ``reference'' cosmological model, and the
quantities without the subscript mean the underlying true values. Since
the reference cosmological model we assume generally differs from the
underlying true cosmology, an apparent geometrical distortion is caused
in the two-dimensional pattern of galaxy clustering. In principle, this distortion
could be measured using only the isotropy of clustering statistics, the
so-called Alcock-Paczynski (AP) test
\cite{AlcockPaczynski:79}, but a more robust measurement of both $D_A(z)$ and
$H(z)$ can be obtained by searching for the ``common'' BAO scales in
the pattern of galaxy clustering, as the standard ruler, in combination
with the CMB constraints
\citep{SeoEisenstein:03,HuHaiman:03}.

We will use the currently standard $\Lambda$CDM model as a guidance for
the parameter dependence of our constraints and as an effective
realistic description of the galaxy clustering. 
To be more quantitative, we assume that 
the redshift-space galaxy power spectrum measured from a hypothetical 
survey realization 
is given in the linear regime as
\begin{eqnarray}
 P^{\rm 2D, obs}_{sW, {\rm ref}}\!(k_{\parallel, {\rm ref}},k_{\perp, {\rm ref}};
  \tau_{33})
  =
\frac{D_{A,{\rm
 ref} }^2 H}
{D_A^2 H_{\rm ref}}
P^{\rm 2D}_{sW}\!\left(k_{\parallel},k_{\perp}; \tau_{33}\right)+P_{\rm sn},
\label{eq:Pg}
\end{eqnarray}
where $P^{\rm 2D, obs}_{sW}$ is the ``observed'' or ``estimated''
power spectrum from a given survey realization, $P^{\rm 2D}_{sW}$ on the right-hand
 side is the true, underlying true power spectrum (Eq.~\ref{eq:p2d_sW}), measured if an observer employs the true 
cosmological model,  and $P_{\rm
sn}$ is a parameter (constant number) to model a possible contamination of a residual shot
noise to the power spectrum measurement. 

To make the parameter forecast,
we employ the method developed in
Ref.~\cite{SeoEisenstein:07,Takadaetal:14,TakadaDore:15}. 
Assuming that the redshift-space power spectrum is measured from a hypothetical survey volume
the Fisher information matrix of model parameters can be computed as
\begin{eqnarray}
F^{\rm galaxy}_{\alpha\beta}&=&\int_{-1}^1\!d\mu\int^{k_{\rm max}}_{k_{\rm
 min}}\!\frac{2\pi k^2\mathrm{d}k}{2(2\pi)^3}
\frac{\partial \ln P^{\rm 2D, obs}_{sW, {\rm ref}}(k,\mu; z_i)}{\partial
 p_\alpha}
\frac{\partial \ln P^{\rm 2D, obs}_{sW, {\rm ref}}(k,\mu; z_i)}{\partial
 p_\beta}
\nonumber\\
&&\hspace{2em}
\times V_{\rm eff}(k; z_i)
\exp\left[-k^2\Sigma_\perp^2 -k^2\mu^2(\Sigma_\parallel^2-\Sigma_\perp^2)\right],
\label{eq:fisher}
\end{eqnarray}
where $\partial P^{\rm 2D, obs}_{sW,{\rm ref}}/\partial p_\alpha$ is the
partial derivative of the galaxy power spectrum (Eq.~\ref{eq:Pg}) with
respect to the $\alpha$-th parameter around the reference cosmological
model. The effective survey volume $V_{\rm eff}$ 
and the Lagrangian
displacement fields $\Sigma_\parallel$ and $\Sigma$ to model the
smearing effect are given
as
\begin{eqnarray}
V_{\rm eff}(k,\mu;z_i)&\equiv&
 \left[\frac{\bar{n}_g(z_i)P^{\rm 2D}_{sW}(k,\mu;z_i)}
{\bar{n}_g(z_i)P^{\rm 2D}_{sW}(k,\mu; z_i)+1}\right]^2V_{\rm
 survey}(z_i),\\
\Sigma_{\perp}(z)&\equiv & c_{\rm rec}D(z)\Sigma_0, \\
\Sigma_{\parallel}(z)&\equiv & c_{\rm rec}D(z)(1+f_g)\Sigma_0.
\label{eq:sigma}
\end{eqnarray}
Here $V_{\rm survey}(z_i)$ is the comoving volume of the redshift slice
centered at $z_i$, $\Sigma_0$ is 
the present-day Lagrangian displacement field, given as
$\Sigma_0=11h^{-1}{\rm Mpc}$ for $\sigma_8=0.8$ \cite{Eisensteinetal:07}, 
and the parameter
$c_{\rm rec}$ is a parameter to model the reconstruction method of the
BAO peaks (see below). 
In
Eq.~(\ref{eq:fisher}), we take the exponential factor of the smearing
effect outside of the derivatives of $P^{\rm 2D, obs}_{sW,{\rm ref}}$. This is equivalent
to marginalizing over uncertainties in $\Sigma_\parallel$ and
$\Sigma_\perp$.
We include the parameter for the large-scale tides for the survey volume, 
i.e. 
$\tau_{33}$ in addition to 
the cosmological parameters, the distances in each redshift slice, and other
nuisance parameters:
\begin{eqnarray}
p_{\alpha}&=&\{\tau_{33},\Omega_{\rm m0}, A_s, n_s, \alpha_s, \Omega_{\rm m0}h^2,
\Omega_{\rm b0}h^2, D_A(z_i), 
H(z_i), b_g(z_i), 
\beta(z_i), P_{\rm
sn}(z_i)  \}, 
\label{eq:parameters}
\end{eqnarray}
where $A_s$, $n_s$ and $\alpha_s$ are parameters of the primordial power
spectrum; $A_s$ is the amplitude of the primordial curvature
perturbation, and $n_s$ and $\alpha_s$ are the spectral tilt and the
running spectral index. The set of cosmological parameters determines
the shape of the linear power spectrum.
For the
$k$-integration, we set $k_{\rm min}=10^{-4}h/{\rm Mpc}$ and $k_{\rm
max}=0.5~h/{\rm Mpc}$, but the exponential factor in
Eq.~(\ref{eq:fisher}) suppresses the information from the nonlinear
scales. The Fisher parameter forecasts depend on the reference
cosmological model for which we assumed the model consistent with the
WMAP 7-year data \cite{WMAP7}. In this paper, we consider a single redshift slice, and
then consider 12 parameters in total in the Fisher analysis.

Furthermore, we assume the BAO reconstruction method in
Ref.~\cite{Eisensteinetal:07}.  Because the large-scale peculiar velocity
field of galaxies in large-scale structure can be inferred from the measured
galaxy distribution, the inferred velocity field allows for pulling back
each galaxy to its position at an earlier epoch and then reconstructing
the galaxy distribution more in the linear regime. As a result, one can
correct to some extent the smearing effect in Eq.~(\ref{eq:fisher}) and
sharpen the BAO peaks in the galaxy power spectrum. 
Padmanabhan et al. \citep{Padmanabhanetal:12} implemented this method to the real data,
SDSS DR7 LRG catalog, and showed that the reconstruction method can
improve the distance error by a factor of 2. The improvement was
equivalent to reducing the nonlinear smoothing scale from $8.1$ to
$\Sigma_{\rm nl}=4.4~h^{-1}{\rm Mpc}$, about a factor of 2 reduce in the
displacement field. 
In the Fisher matrix calculation, we used $c_{\rm rec}=0.5$ as a default choice 
\cite{Padmanabhanetal:12}.

In the following forecast, we assume the BAO experiment combined with
the CMB constraints expected from the Planck satellite:
\begin{equation}
\bm{F}=\bm{F}^{\rm CMB} + \bm{F}^{\rm galaxy},
\end{equation}
where $\bm{F}_{\rm CMB}$ is the Fisher matrix for the CMB measurements.
We employ the method in Ref.~\cite{Takadaetal:14} to compute the CMB
Fisher matrix, where we assumed the standard $\Lambda$CDM model for the
physics prior to recombination that determines the sound horizon scale
or the BAO scale.

\subsection{Results}

As a working example, we consider a hypothetical survey that is characterized by the central redshift 
$z=0.5$, the comoving volume $V=1~({\rm Gpc}/h)^3$, the mean number density of galaxies 
$\bar{n}_g=10^{-3}~(h/{\rm Mpc})^3$ and linear bias parameter $b=2$, respectively.
For simplicity we consider a single redshift slice. 
In reality, when a galaxy redshift survey probes galaxies over a wide range of redshifts, one 
can use the clustering analysis in multiple redshift slices and then combine their cosmological information. 

\begin{figure}
\centering \includegraphics[width=18cm]{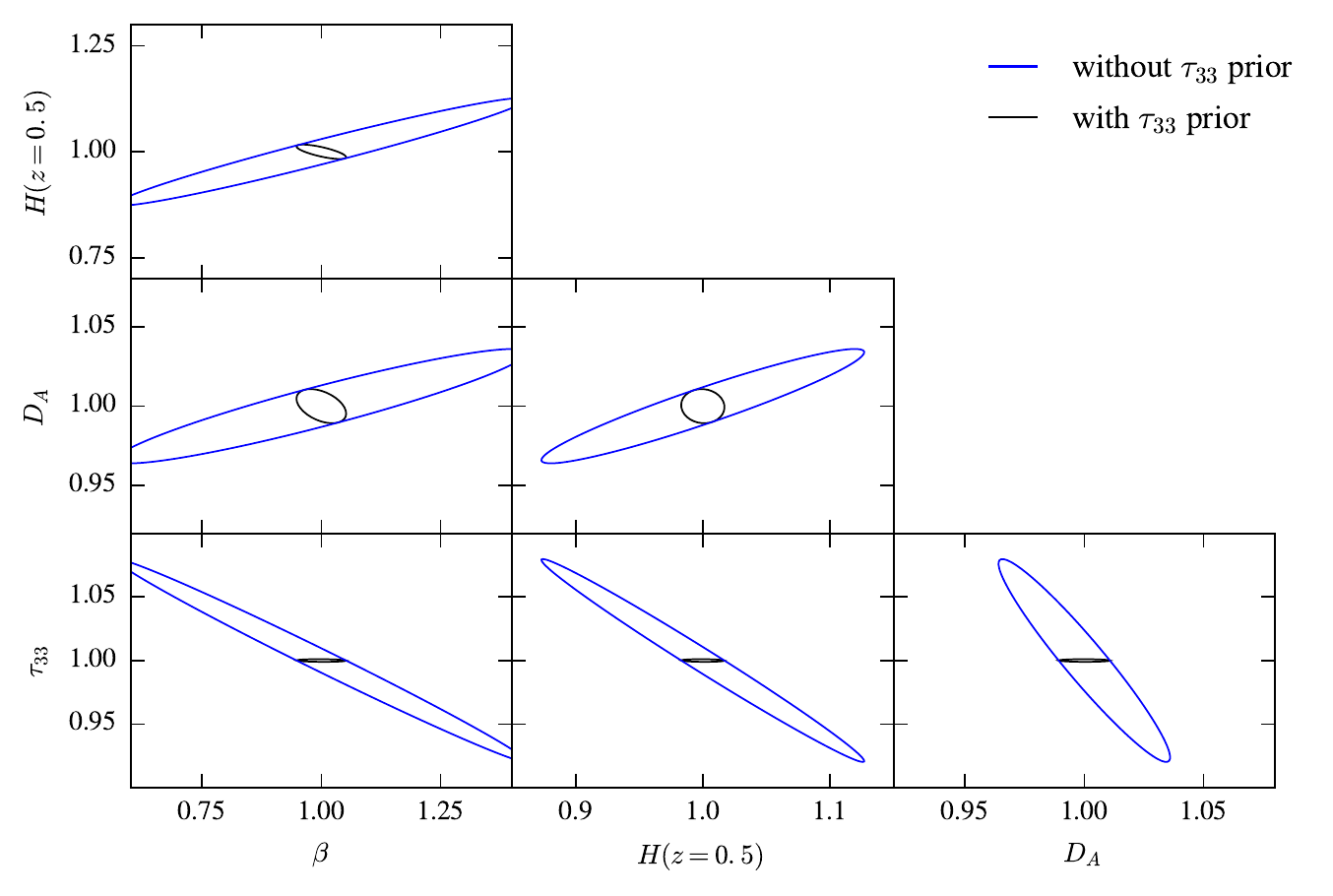}
\caption{68\% CL error ellipse for the parameters, 
$\tau_{33}$, $D_A$, $H$ and $\beta$, including marginalization over other parameters in the Fisher analysis (see Section~\ref{sec:Fisher} for details). The inner black contour in each panel 
shows the result when $\sigma_{\tau_{33}}=1.04\times 10^{-3}$ is employed as the $\tau_{33}$ prior, which is taken from 
the rms value expected for the $\Lambda$CDM model and the assumed galaxy survey that is characterized by $V=1~({\rm Gpc}/h)^3$, 
$\bar{n}_{\rm g}=10^{-3}~(h/{\rm Mpc})^3$ and $b=2$.
 }
\label{fig:Fisher_full}

\end{figure}
\begin{figure}
\begin{tabular}{c}

	\begin{minipage}{0.33\hsize}
		\begin{center}
			\includegraphics[width=6cm]{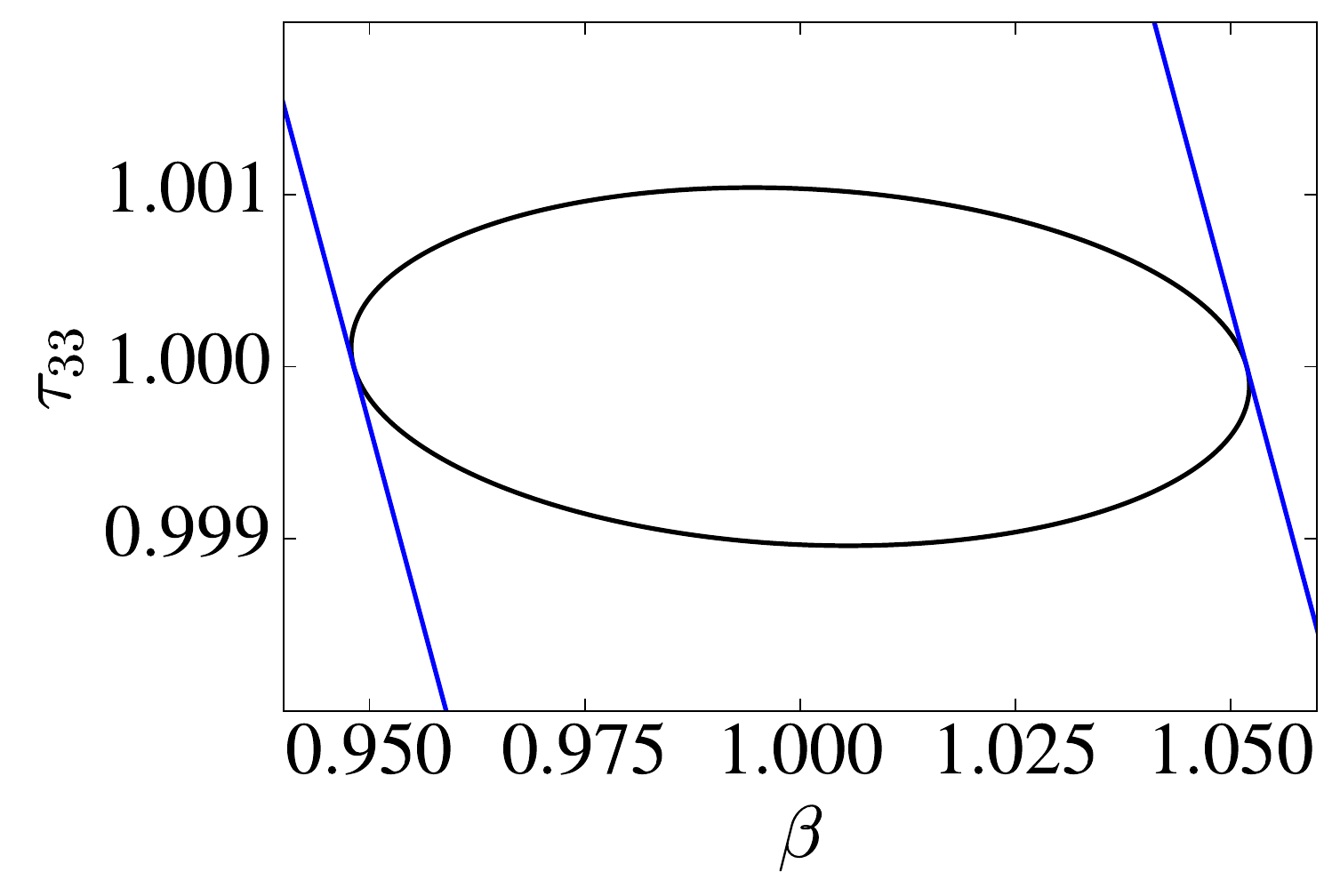}
			Fisher ellipse for $\tau_{33}$ and $\beta$
		\end{center}
	\end{minipage}

	\begin{minipage}{0.33\hsize}
		\begin{center}
			\includegraphics[width=6cm]{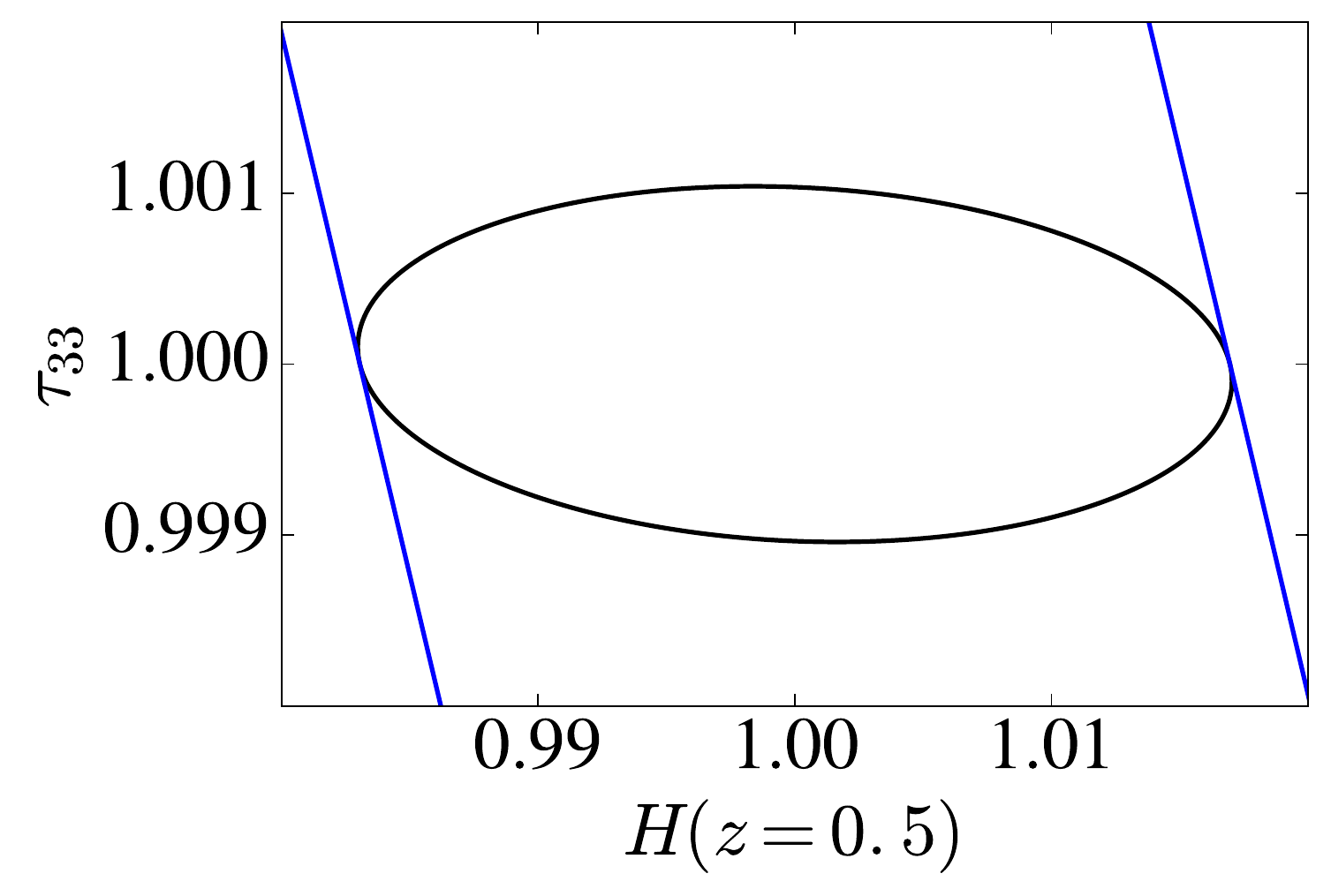}
			Fisher ellipse for $\tau_{33}$ and $H$
		\end{center}
	\end{minipage}

	\begin{minipage}{0.33\hsize}
		\begin{center}
			\includegraphics[width=6cm]{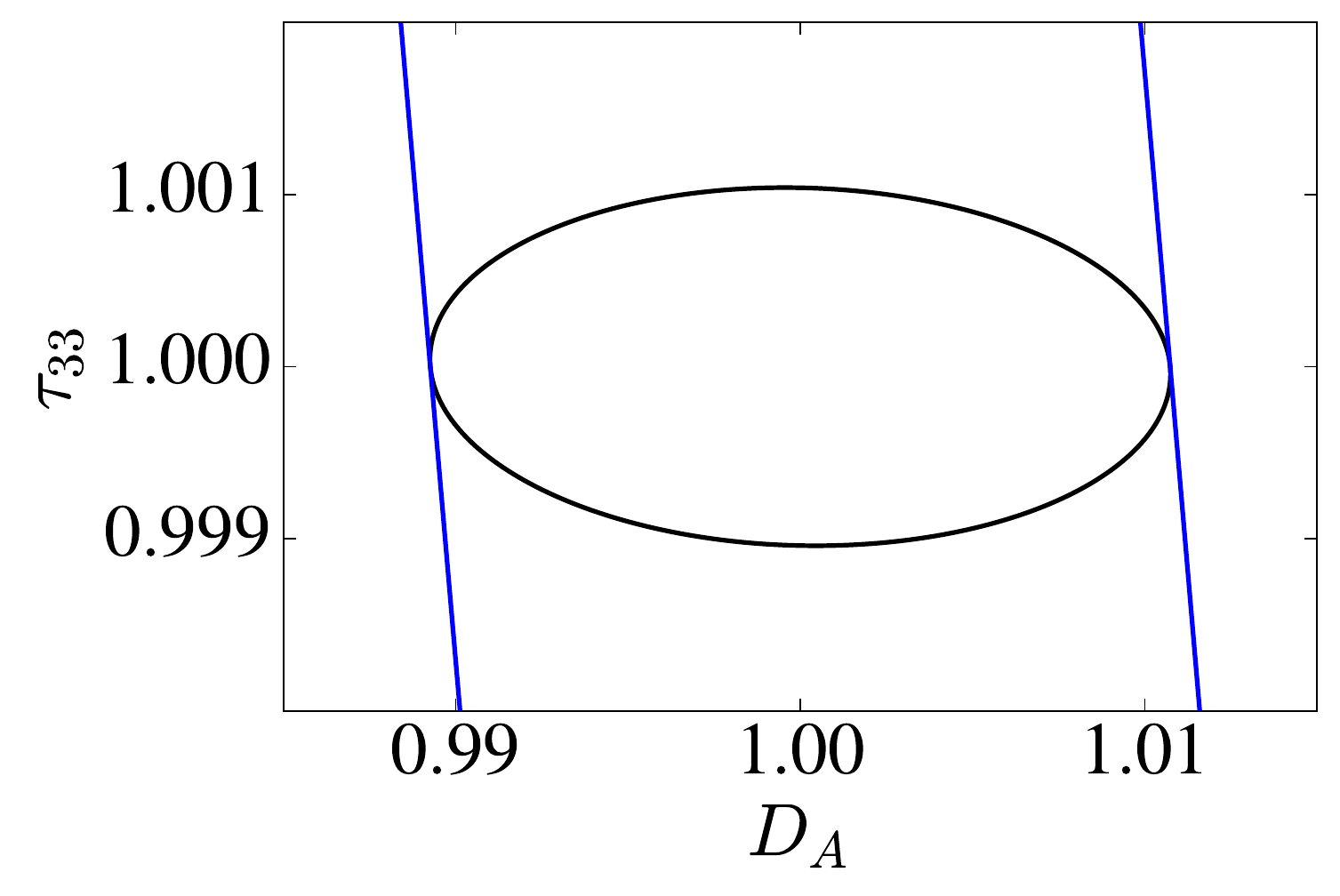}
			Fisher ellipse for $\tau_{33}$ and $D_A$
		\end{center}
	\end{minipage}

\end{tabular}
\caption{A zoom-in version of Fig.~\ref{fig:Fisher_full}, around the fiducial model for the Fisher analysis.}
\label{fig:Fisher_zoom}
\end{figure}

\begin{figure}
\begin{center}
	\includegraphics[width=18cm]{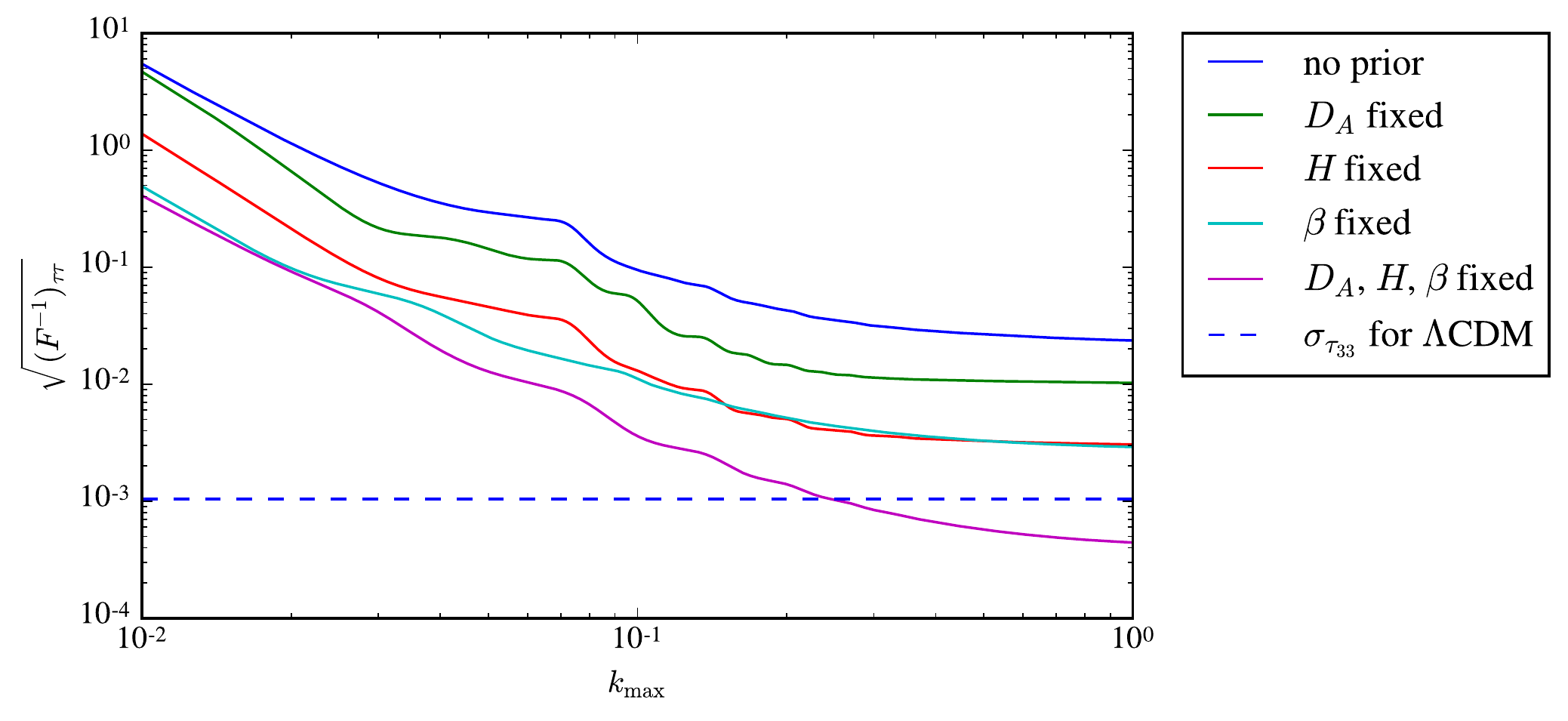}
	\caption{The marginalized error on the estimation of $\tau_{33}$, 
	$\sqrt{(F^{-1})_{\tau\tau}}$, as a function of the maximum wavenumber $k_{\rm max}$ 
	up to which the redshift-space power spectrum
	information is included in the Fisher analysis (see text for the details). The different solid curves show the results when any prior on other 
	parameters ($D_A, H$ and $\beta$) are not employed or when some or all the parameters are fixed to their values for the $\Lambda$CDM model. 
	The horizontal dashed curve is the rms value, $\sigma_{\tau_{33}}$, expected for the $\Lambda$CDM model and the survey volume. 
	Note that we did not impose any prior on other parameters (Eq.~\ref{eq:parameters}), although the CMB information is added. 
	}
	\label{fig:F_inv_tau}
\end{center}
\end{figure}

In Fig.~\ref{fig:Fisher_full} we show the marginalized 68\% CL error contours in each of two-dimensional sub-space that include either two of
 the large-scale tidal parameter, $\tau_{33}$, the
distance parameters, $D_A$ or $H$, or the RSD parameter $\beta$, where the contours include marginalization over other 
parameters. Note that $\tau_{33}$ has little degeneracy with other parameters. More quantitatively, 
 the cross-correlation coefficients defined as $c_{ij}=(F^{-1})_{ij}/\sqrt{(F^{-1})_{ii}(F^{-1})_{jj}}$ with $i={\tau_{33}}$, after the CMB Fisher matrix is added, 
is almost unity for either one of
these three parameters is taken for $j$, while the cross-coefficients are smaller for other parameters, less than $\mathcal{O}(0.2)$.
The contours in each panel of Fig.~\ref{fig:Fisher_full} show how an uncertainty in $\tau_{33}$ causes a degeneracy with estimation of other parameter. 
Since the large-scale tides cause apparent anisotropies in the observed clustering of galaxies as the radial AP anisotropy and the RSD effect do, allowing $\tau_{\rm 33}$ to freely vary in the parameter estimation causes significant degeneracies with $\beta$ and $H$. The degeneracy between 
$\tau_{33}$ and $D_A$ arises from the trace-less nature of $\tau_{ij}$; changing $\tau_{33}$ leads to a change in $\tau_{11}+\tau_{22}(=-\tau_{33})$ and  therefore  causes an apparent distortion in the $k_{\perp}$-direction, which mimics the cosmological distortion due to a change in $D_A$.

However, if adding the prior on $\tau_{33}$ assuming the $\Lambda$CDM model, i.e. $\tau_{\rm 33}=0$ for the expectation value and 
$\sigma_{\tau_{33}}=1.04\times 10^{-3}$ for the rms value for $V=1~({\rm Gpc}/h)^3$, it lifts the degeneracies, recovering a high-precision 
measurement for each cosmological parameter. Fig.~\ref{fig:Fisher_zoom} shows a zoom-in version of the contours around the central value (the input model in the Fisher analysis), and shows that the prior of $\tau_{33}$ efficiently breaks the parameter degeneracies. In particular, even if an actual value of $\tau_{33}$ in a given survey realization is off from zero by more than a few $\sigma_{\tau_{33}}$, it does not seem to cause a significant bias in 
the parameters.

Nevertheless it is interesting to ask whether a measurement of redshift-shift power spectrum of galaxies can be used to constrain the large-scale tides,
 $\tau_{33}$, rather than employing the prior, if one can include the information up to the larger $k$ beyond the weakly nonlinear regime. To address this possibility, we need to know the response of the redshift-space power spectrum to the tides, $\partial P_{s}(\bk)/\partial \tau_{33}$, in the nonlinear regime where  the perturbation theory breaks down. To estimate the response function in the nonlinear regime requires to, e.g. use a separate universe simulation where the large-scale tidal effect is included in the background expansion, similarly to the method used for estimating the response for the mean density modulation, $\partial P_s/\partial \delta_\br$, in Refs.~\cite{Lietal:14a,Lietal:14b,Wagneretal:15,Baldaufetal:16}. This is beyond the scope of this paper, so here we simply assume that the response function derived using the perturbation theory holds in the nonlinear regime. This would be conservative, because the response is likely to be amplified in the nonlinear regime as shown in Ref.~\cite{Lietal:14a}. Furthermore, to include the effect of the large-scale tides up to the nonlinear regime, we set $\Sigma=0$ for the BAO smearing factor in the Fisher analysis. In practice, the smearing factor also depends on nonlinear structure formation, and therefore would depend on $\tau_{\rm 33}$. This is a simplified assumption, but we believe that the following result gives a rough estimation of the genuine effect. 
Fig.~\ref{fig:F_inv_tau} shows how an accuracy of the $\tau_{33}$ estimation is improved when including the redshift-space power spectrum information up to a given maximum wavenumber $k_{\rm max}$.
Without any prior, $\tau_{33}$ is estimated to about $1\%$ accuracy for a survey volume of $V=1~({\rm Gpc}/h)^3$. 
When fixing other parameters to their values for $\Lambda$CDM model, the accuracy of $\tau_{33}$ is dramatically improved. In particular, 
when all the distortion parameters, $D_A, H$ and $\beta$, are fixed, the $\tau_{33}$ parameter could be determined to an accuracy better than 
the rms for $\Lambda$CDM model, if the redshift-space power spectrum information is included up to $k_{\rm max}\gtrsim 0.25~h/{\rm Mpc}$.
This result implies that the anisotropic clustering information 
in such a nonlinear regime could be used to infer the large-scale tides for a given survey realization. 

\section{Discussion}
\label{sec:discussion}

In this paper, using the standard perturbation theory, we derived
the response functions of the redshift-space power spectrum to the super-survey modes,
both the isotropic component, $\partial P_s(\bk)/\partial \delta_{\rm b}$, and the anisotropic components, $\partial P_s(\bk)/\partial \tau_{ij}$.
Since a given survey realization is generally embedded in the presence of super-survey modes, $\delta_{\rm b}$ and $\tau_{ij}$, that are not direct observables in a finite-volume survey, it is important to take into account 
the response functions which describe how the super-survey modes cause a modulation in the redshift-space power spectrum measured in the survey volume, compared with the ensemble average expectation for an infinite volume. There are two effects. First, the presence of 
the super-survey modes changes the growth of small-scale fluctuations via the nonlinear mode coupling. Secondly, it
causes a dilation effect, the modulation of a short distance scale due to the change of the local expansion factor in the finite volume region. 
In particular we showed that the large-scale tides, $\tau_{ij}$, cause an apparent anisotropic clustering in the redshift-space power spectrum, where the effect has directional dependence determined by an alignment of the large-scale tides, the directions of 
 small-scale  modes, and the line-of-sight direction. 
This effect mimics an anisotropic clustering due to the 
redshift-space distortion effect of the small-scale peculiar velocities of galaxies as well as 
the apparent cosmological distortion caused by the use of an incorrect cosmological model in the clustering analysis. 

To assess a possible impact of $\tau_{ij}$ on parameter estimation from a measurement of the redshift-space power spectrum in a given survey realization, we 
used the Fisher information matrix formalism. To do this, we treated the effect of $\tau_{ij}$ as a ``signal'' rather than 
 an additional source of statistical errors in the redshift-space power spectrum measurement, because 
it causes a modulation in the measured power spectrum as do cosmological parameters around the true model: $P_s(\bk;\tau_{ij})= P_s(\bk)+\tau_{ij}\partial P_s(\bk)/\partial \tau_{ij}$, where the tensor $\tau_{ij}$ takes particular values in a given survey realization. Thus as long as an accurate model of the response function is given as a function of cosmological models, it would be straightforward to include the effect in parameter estimation. 
In this paper, we considered the two-dimensional redshift-space power spectrum, $P^{\rm 2D}_{s}(k_\perp,k_\parallel;\tau_{33})$ as an observable, which is obtained from the azimuthal angle average of the redshift-space power spectrum estimator in the two-dimensional plane perpendicular to the line-of-sight direction under the distant observer approximation. In this case, the effects of the large-scale tides are modeled by a single quantity, $\tau_{33}$, the line-of-sight component of the tides. 
We showed that, if allowing $\tau_{33}$ to freely vary, it causes a significant degradation in the parameters, 
$D_A, H$ and $\beta$, due to
almost perfect degeneracies between $\tau_{33}$ and the parameters in the power spectrum. If one adopts a prior on $\tau_{33}$ assuming 
the rms expected for a $\Lambda$CDM model, it efficiently lifts the parameter degeneracies and restores an accuracy 
of the cosmological parameters that are expected for a galaxy survey without the super-survey mode. 
Thus the impact of the large-scale tides on the redshift-space power spectrum is not as large as the impact of the large-scale density contrast, 
$\delta_\br$, on a real-space power spectrum such as the weak lensing power spectrum \citep{TakadaHu:13,Lietal:14a}, as long as the large-scale tides obey the $\Lambda$CDM expectation. 
The reason for this less-significant impact is partly because the statistical uncertainty in a measurement of the quadrupole power spectrum,
which is the lowest-order observable to extract the redshift-space distortion, is dominated by the statistical uncertainty in the 
monopole power spectrum measurement \cite{Akitsuetal:17}.

We have also addressed whether a measurement of redshift-space power spectrum can be used to {\em constrain}  
$\tau_{33}$ in the survey realization, rather than treating $\tau_{33}$ as a nuisance parameter. Because the presence of $\tau_{33}$ causes 
a mode-coupling with all the small-scale fluctuations, we showed that $\tau_{33}$ can be well constrained at an accuracy better than the rms 
for a $\Lambda$CDM model, if we can use the redshift-space power spectrum information up to small scales, $k_{\rm max}\gtrsim 0.25~{h/{\rm Mpc}}$ and if 
the cosmological parameters including $D_A, H$ and $\beta$ are sufficiently well constrained, e.g. by other cosmological probes.  
This is an interesting possibility, because the method gives an access to such large-scale tides from a measurement of the small-scale fluctuations, and 
the large-scale 
tides would contain the information on physics in the early universe, e.g. statistical anisotropies arising from the inflation physics \cite{JeongKamionkowski:12} or a large-scale anisotropy due to the super-curvature fluctuation \cite{Byrnesetal:16}.

However, there are several limitations in the results shown in this paper. First, 
we used the perturbation theory prediction for the response function of redshift-space power spectrum in the Fisher analysis, which breaks down in the deeply nonlinear regime. In order to realize the genuine functional form of the response function in the nonlinear regime, 
we need to use $N$-body simulations of large-scale structure formation and then study a coupling of the large-scale tidal modes with small-scale Fourier modes. For doing this, a ``separate universe simulation'' method would be powerful, where the large-scale modes are absorbed into the background expansion. It was shown that this method works very well to model the response function to the large-scale density contrast, $\partial P(k)/\partial 
\delta_\br$ \cite{Lietal:14a,Lietal:14b,Wagneretal:15,Lietal:16,Baldaufetal:16,Huetal16}. To extend this method, one can adopt an anisotropic expansion background to model the effect of the anisotropic super-survey mode, $\tau_{ij}$, and then run  an $N$-body simulation onto the modified background \citep[e.g., see Ref.][for the related discussion]{IpSchmidt:17}. If this separate universe simulation 
for $\tau_{ij}$ is developed, one can study various effects of $\tau_{ij}$ on nonlinear structures; the non-local bias of halos \citep{Chanetal:12,Saitoetal:14}, the correlation between $\tau_{ij}$ and shapes of halos \cite{Chisarietal:16,Okumuraetal:17}, and so on. This would be very interesting, and is our future work. 

Another limitation of this paper is we used the redshift-space power spectrum, $P_s(k_\perp,k_\parallel)$,
which is given as a function of two wavenumber variables such as $k_\perp$ and $k_\parallel$. 
Since the principle axes directions of $\tau_{ij}$ have nothing with the line-of-sight direction, the effect of $\tau_{ij}$ generally violates an statistical isotropy in the two-dimensional plane perpendicular to the line-of-sight direction. Hence, in order to fully extract 
the three-dimensional information on the tensor $\tau_{ij}$, one needs to use
 the redshift-space power spectrum given as a function of the three-dimensional vector, $P_s(\bk)$, without employing the angle average in the perpendicular plane as usually done in the standard method. Alternatively
one can use a more general expansion of the redshift-space power spectrum, e.g. the bipolar spherical harmonics (BipoSH) decomposition \cite{Shiraishietal:17}. It would be interesting to study how the full information on $\tau_{ij}$ can be extracted by using the BipSH decomposition. 

\smallskip{\em Acknowledgments.--} We thank Yin Li, Takahiro Nishimichi, Fabian Schmidt,
Maresuke Shiraishi and Naonori Sugiyama for useful discussion, and we
also thank to YITP, Kyoto University for their warm hospitality. KA is
supported by the Advanced Leading Graduate Course for Photon Science at
the University of Tokyo. MT is supported by World Premier International
Research Center Initiative (WPI Initiative), MEXT, Japan, by the FIRST
program ``Subaru Measurements of Images and Redshifts (SuMIRe)'', CSTP,
Japan. MT is supported by Grant-in-Aid for Scientific Research from the
JSPS Promotion of Science (No.~23340061, 26610058, and 15H05893), MEXT Grant-in-Aid
for Scientific Research on Innovative Areas (No.~15K21733,
15H05892) and by JSPS Program for Advancing Strategic International
Networks to Accelerate the Circulation of Talented Researchers.

\appendix
\section{Multipole power spectrum in the redshift-space power spectrum}
\label{app:multipole}

Here, we show the multipole expansion of 2D power spectrum in the redshift-space.
The multipole power spectrum are defined as 
\begin{equation}
P_{s}^{\ell}(k;\delta_b, \tau_{33}) \equiv (2\ell +1 ) \int_{-1}^{1} \frac{d\mu}{2}\; P^{2D}_{sW}(k, \mu; \delta_b, \tau_{33}) \mathcal{L}_{\ell} (\mu),
\label{eq:multi_def}
\end{equation}
where $\mathcal{L}_{\ell}(\mu)$ is the Legendre polynomial.
Making the use of Eq.~(\ref{eq:dpsddb}) and Eq.~(\ref{eq:p2d_sW}), the multipole power spectra in the redshift-space can be calculated as
 {\allowdisplaybreaks
\begin{eqnarray}
P^{\ell=0}_s &=& \left[ b^2+\frac{2}{3}bf+\frac{1}{5}f^2\right] P(k) \nonumber \\
&& + \delta_b\left[ \left\{ \frac{94}{21}b^2+\frac{52}{21}bf+2b^2f +\frac{62}{105}f^2+\frac{4}{3}bf^2+\frac{26}{105}f^3 \right\}P(k) \right.\nonumber \\
&& \hspace{1cm}- \left. \left\{ \frac{2}{3}b^2+2b^2f +\frac{4}{9}bf+\frac{2}{9}b^2f+\frac{2}{15}f^2+\frac{4}{15}fb^2+\frac{2}{21}f^3 \right\} 
 \frac{\partial P(k)}{\partial \ln{k}} \right] \nonumber 
\\
&& + \tau_{33}\left[ \left\{ 2b^2f+ \frac{32}{35}bf +\frac{128}{245}f^2+\frac{8}{5}bf^2+\frac{26}{35}f^3 \right\}P(k) \right.\nonumber \\
&& \hspace{1cm}- \left. \left\{ \frac{8}{15}bf+\frac{2}{3}b^2f+\frac{8}{35}f^2+\frac{4}{5}bf^2+\frac{2}{7}f^3\right\} 
 \frac{\partial P(k)}{\partial \ln{k}} \right] \\
P^{\ell=2}_s &=& \left[ \frac{4}{3}bf+\frac{4}{7}f^2\right] P(k) \nonumber \\
&& + \delta_b\left[ \left\{ \frac{8}{15}b^2f+\frac{104}{105}bf+\frac{16}{21}bf^2 +\frac{248}{735}f^2+\frac{8}{45}f^3\right\}P(k) \right. \nonumber \\
&& \hspace{1cm}- \left. \left\{ \frac{4}{45}b^2+\frac{8}{45}bf+\frac{16}{105}bf^2 +\frac{8}{105}f^2 +\frac{4}{63}f^3\right\} 
 \frac{\partial P(k)}{\partial \ln{k}} \right] \nonumber 
\\
&& + \tau_{33}\left[ \left\{ \frac{16}{35}b^2 +\frac{176}{245}bf+\frac{96}{245}f^2+\frac{32}{35}bf^2+\frac{8}{15}f^3 \right\}P(k) \right. \nonumber \\
&& \hspace{1cm}- \left. \left\{ \frac{2}{5}b^2+\frac{44}{105}bf+\frac{4}{15}b^2f+\frac{6}{35}f^2+\frac{16}{35}bf^2 +\frac{4}{21}f^3\right\} 
 \frac{\partial P(k)}{\partial \ln{k}} \right]\\
P^{\ell=4}_s &=& \frac{8}{35}f^2 P(k) \nonumber  \\
&& + \delta_b\left[ \left\{ \frac{496}{6615}f^2+\frac{32}{189}bf^2+\frac{112}{1485}f^3\right\}P(k) \right. \nonumber \\
&& \hspace{1cm}- \left. \left\{ \frac{16}{945}f^2+\frac{32}{945}bf^2+\frac{16}{693}f^3\right\} 
 \frac{\partial P(k)}{\partial \ln{k}} \right] \nonumber 
\\
&& + \tau_{33}\left[ \left\{ \frac{64}{245}bf +\frac{4352}{24255}f^2+\frac{64}{315}bf^2+\frac{112}{495}f^3 \right\}P(k) \right. \nonumber \\
&& \hspace{1cm}- \left. \left\{ \frac{16}{105}bf+\frac{272}{3465}f^2+\frac{32}{315}bf^2+\frac{16}{231}f^3 \right\} 
 \frac{\partial P(k)}{\partial \ln{k}} \right] \\
P^{\ell=6}_s &=&  \delta_b\left[ \frac{128}{9009}f^3P(k) - \frac{32}{9009}f^3 \frac{\partial P(k)}{\partial \ln{k}} \right] \nonumber  \\
&& + \tau_{33}\left[ \left\{ \frac{256}{7007}f^2 +\frac{128}{3003}f^3 \right\}P(k)
- \left\{ \frac{16}{1001}f^2+\frac{32}{3003}f^3 \right\} \frac{\partial P(k)}{\partial \ln{k}} \right]
\end{eqnarray}
}
and the higher-multipole spectra with $\ell\ge 8$ vanish.

\bibliography{sst}

\begin{thebibliography}{62}%
\makeatletter
\providecommand \@ifxundefined [1]{%
 \@ifx{#1\undefined}
}%
\providecommand \@ifnum [1]{%
 \ifnum #1\expandafter \@firstoftwo
 \else \expandafter \@secondoftwo
 \fi
}%
\providecommand \@ifx [1]{%
 \ifx #1\expandafter \@firstoftwo
 \else \expandafter \@secondoftwo
 \fi
}%
\providecommand \natexlab [1]{#1}%
\providecommand \enquote  [1]{``#1''}%
\providecommand \bibnamefont  [1]{#1}%
\providecommand \bibfnamefont [1]{#1}%
\providecommand \citenamefont [1]{#1}%
\providecommand \href@noop [0]{\@secondoftwo}%
\providecommand \href [0]{\begingroup \@sanitize@url \@href}%
\providecommand \@href[1]{\@@startlink{#1}\@@href}%
\providecommand \@@href[1]{\endgroup#1\@@endlink}%
\providecommand \@sanitize@url [0]{\catcode `\\12\catcode `\$12\catcode
  `\&12\catcode `\#12\catcode `\^12\catcode `\_12\catcode `\%12\relax}%
\providecommand \@@startlink[1]{}%
\providecommand \@@endlink[0]{}%
\providecommand \url  [0]{\begingroup\@sanitize@url \@url }%
\providecommand \@url [1]{\endgroup\@href {#1}{\urlprefix }}%
\providecommand \urlprefix  [0]{URL }%
\providecommand \Eprint [0]{\href }%
\providecommand \doibase [0]{http://dx.doi.org/}%
\providecommand \selectlanguage [0]{\@gobble}%
\providecommand \bibinfo  [0]{\@secondoftwo}%
\providecommand \bibfield  [0]{\@secondoftwo}%
\providecommand \translation [1]{[#1]}%
\providecommand \BibitemOpen [0]{}%
\providecommand \bibitemStop [0]{}%
\providecommand \bibitemNoStop [0]{.\EOS\space}%
\providecommand \EOS [0]{\spacefactor3000\relax}%
\providecommand \BibitemShut  [1]{\csname bibitem#1\endcsname}%
\let\auto@bib@innerbib\@empty
\bibitem [{\citenamefont {{Takada}}\ \emph {et~al.}(2006)\citenamefont
  {{Takada}}, \citenamefont {{Komatsu}},\ and\ \citenamefont
  {{Futamase}}}]{Takadaetal:06}%
  \BibitemOpen
  \bibfield  {author} {\bibinfo {author} {\bibfnamefont {M.}~\bibnamefont
  {{Takada}}}, \bibinfo {author} {\bibfnamefont {E.}~\bibnamefont {{Komatsu}}},
  \ and\ \bibinfo {author} {\bibfnamefont {T.}~\bibnamefont {{Futamase}}},\
  }\href {\doibase 10.1103/PhysRevD.73.083520} {\bibfield  {journal} {\bibinfo
  {journal} {\prd}\ }\textbf {\bibinfo {volume} {73}},\ \bibinfo {pages}
  {083520} (\bibinfo {year} {2006})},\ \Eprint
  {http://arxiv.org/abs/arXiv:astro-ph/0512374} {arXiv:astro-ph/0512374}
  \BibitemShut {NoStop}%
\bibitem [{\citenamefont {{Takada}}\ \emph {et~al.}(2014)\citenamefont
  {{Takada}}, \citenamefont {{Ellis}}, \citenamefont {{Chiba}}, \citenamefont
  {{Greene}}, \citenamefont {{Aihara}}, \citenamefont {{Arimoto}},
  \citenamefont {{Bundy}}, \citenamefont {{Cohen}}, \citenamefont {{Dor{\'e}}},
  \citenamefont {{Graves}}, \citenamefont {{Gunn}}, \citenamefont {{Heckman}},
  \citenamefont {{Hirata}}, \citenamefont {{Ho}}, \citenamefont {{Kneib}},
  \citenamefont {{F{\`e}vre}}, \citenamefont {{Lin}}, \citenamefont {{More}},
  \citenamefont {{Murayama}}, \citenamefont {{Nagao}}, \citenamefont {{Ouchi}},
  \citenamefont {{Seiffert}}, \citenamefont {{Silverman}}, \citenamefont
  {{Sodr{\'e}}}, \citenamefont {{Spergel}}, \citenamefont {{Strauss}},
  \citenamefont {{Sugai}}, \citenamefont {{Suto}}, \citenamefont {{Takami}},\
  and\ \citenamefont {{Wyse}}}]{Takadaetal:14}%
  \BibitemOpen
  \bibfield  {author} {\bibinfo {author} {\bibfnamefont {M.}~\bibnamefont
  {{Takada}}}, \bibinfo {author} {\bibfnamefont {R.~S.}\ \bibnamefont
  {{Ellis}}}, \bibinfo {author} {\bibfnamefont {M.}~\bibnamefont {{Chiba}}},
  \bibinfo {author} {\bibfnamefont {J.~E.}\ \bibnamefont {{Greene}}}, \bibinfo
  {author} {\bibfnamefont {H.}~\bibnamefont {{Aihara}}}, \bibinfo {author}
  {\bibfnamefont {N.}~\bibnamefont {{Arimoto}}}, \bibinfo {author}
  {\bibfnamefont {K.}~\bibnamefont {{Bundy}}}, \bibinfo {author} {\bibfnamefont
  {J.}~\bibnamefont {{Cohen}}}, \bibinfo {author} {\bibfnamefont
  {O.}~\bibnamefont {{Dor{\'e}}}}, \bibinfo {author} {\bibfnamefont
  {G.}~\bibnamefont {{Graves}}}, \bibinfo {author} {\bibfnamefont {J.~E.}\
  \bibnamefont {{Gunn}}}, \bibinfo {author} {\bibfnamefont {T.}~\bibnamefont
  {{Heckman}}}, \bibinfo {author} {\bibfnamefont {C.~M.}\ \bibnamefont
  {{Hirata}}}, \bibinfo {author} {\bibfnamefont {P.}~\bibnamefont {{Ho}}},
  \bibinfo {author} {\bibfnamefont {J.-P.}\ \bibnamefont {{Kneib}}}, \bibinfo
  {author} {\bibfnamefont {O.~L.}\ \bibnamefont {{F{\`e}vre}}}, \bibinfo
  {author} {\bibfnamefont {L.}~\bibnamefont {{Lin}}}, \bibinfo {author}
  {\bibfnamefont {S.}~\bibnamefont {{More}}}, \bibinfo {author} {\bibfnamefont
  {H.}~\bibnamefont {{Murayama}}}, \bibinfo {author} {\bibfnamefont
  {T.}~\bibnamefont {{Nagao}}}, \bibinfo {author} {\bibfnamefont
  {M.}~\bibnamefont {{Ouchi}}}, \bibinfo {author} {\bibfnamefont
  {M.}~\bibnamefont {{Seiffert}}}, \bibinfo {author} {\bibfnamefont {J.~D.}\
  \bibnamefont {{Silverman}}}, \bibinfo {author} {\bibfnamefont
  {L.}~\bibnamefont {{Sodr{\'e}}}}, \bibinfo {author} {\bibfnamefont {D.~N.}\
  \bibnamefont {{Spergel}}}, \bibinfo {author} {\bibfnamefont {M.~A.}\
  \bibnamefont {{Strauss}}}, \bibinfo {author} {\bibfnamefont {H.}~\bibnamefont
  {{Sugai}}}, \bibinfo {author} {\bibfnamefont {Y.}~\bibnamefont {{Suto}}},
  \bibinfo {author} {\bibfnamefont {H.}~\bibnamefont {{Takami}}}, \ and\
  \bibinfo {author} {\bibfnamefont {R.}~\bibnamefont {{Wyse}}},\ }\href
  {\doibase 10.1093/pasj/pst019} {\bibfield  {journal} {\bibinfo  {journal}
  {\pasj}\ }\textbf {\bibinfo {volume} {66}},\ \bibinfo {eid} {R1} (\bibinfo
  {year} {2014})},\ \Eprint {http://arxiv.org/abs/1206.0737} {arXiv:1206.0737}
  \BibitemShut {NoStop}%
\bibitem [{\citenamefont {{Takada}}\ and\ \citenamefont
  {{Dor{\'e}}}(2015)}]{TakadaDore:15}%
  \BibitemOpen
  \bibfield  {author} {\bibinfo {author} {\bibfnamefont {M.}~\bibnamefont
  {{Takada}}}\ and\ \bibinfo {author} {\bibfnamefont {O.}~\bibnamefont
  {{Dor{\'e}}}},\ }\href {\doibase 10.1103/PhysRevD.92.123518} {\bibfield
  {journal} {\bibinfo  {journal} {\prd}\ }\textbf {\bibinfo {volume} {92}},\
  \bibinfo {eid} {123518} (\bibinfo {year} {2015})},\ \Eprint
  {http://arxiv.org/abs/1508.02469} {arXiv:1508.02469} \BibitemShut {NoStop}%
\bibitem [{\citenamefont {{Alam}}\ \emph {et~al.}(2017)\citenamefont {{Alam}},
  \citenamefont {{Ata}}, \citenamefont {{Bailey}}, \citenamefont {{Beutler}},
  \citenamefont {{Bizyaev}}, \citenamefont {{Blazek}}, \citenamefont
  {{Bolton}}, \citenamefont {{Brownstein}}, \citenamefont {{Burden}},
  \citenamefont {{Chuang}}, \citenamefont {{Comparat}}, \citenamefont
  {{Cuesta}}, \citenamefont {{Dawson}}, \citenamefont {{Eisenstein}},
  \citenamefont {{Escoffier}}, \citenamefont {{Gil-Mar{\'{\i}}n}},
  \citenamefont {{Grieb}}, \citenamefont {{Hand}}, \citenamefont {{Ho}},
  \citenamefont {{Kinemuchi}}, \citenamefont {{Kirkby}}, \citenamefont
  {{Kitaura}}, \citenamefont {{Malanushenko}}, \citenamefont {{Malanushenko}},
  \citenamefont {{Maraston}}, \citenamefont {{McBride}}, \citenamefont
  {{Nichol}}, \citenamefont {{Olmstead}}, \citenamefont {{Oravetz}},
  \citenamefont {{Padmanabhan}}, \citenamefont {{Palanque-Delabrouille}},
  \citenamefont {{Pan}}, \citenamefont {{Pellejero-Ibanez}}, \citenamefont
  {{Percival}}, \citenamefont {{Petitjean}}, \citenamefont {{Prada}},
  \citenamefont {{Price-Whelan}}, \citenamefont {{Reid}}, \citenamefont
  {{Rodr{\'{\i}}guez-Torres}}, \citenamefont {{Roe}}, \citenamefont {{Ross}},
  \citenamefont {{Ross}}, \citenamefont {{Rossi}}, \citenamefont
  {{Rubi{\~n}o-Mart{\'{\i}}n}}, \citenamefont {{Saito}}, \citenamefont
  {{Salazar-Albornoz}}, \citenamefont {{Samushia}}, \citenamefont
  {{S{\'a}nchez}}, \citenamefont {{Satpathy}}, \citenamefont {{Schlegel}},
  \citenamefont {{Schneider}}, \citenamefont {{Sc{\'o}ccola}}, \citenamefont
  {{Seo}}, \citenamefont {{Sheldon}}, \citenamefont {{Simmons}}, \citenamefont
  {{Slosar}}, \citenamefont {{Strauss}}, \citenamefont {{Swanson}},
  \citenamefont {{Thomas}}, \citenamefont {{Tinker}}, \citenamefont
  {{Tojeiro}}, \citenamefont {{Maga{\~n}a}}, \citenamefont {{Vazquez}},
  \citenamefont {{Verde}}, \citenamefont {{Wake}}, \citenamefont {{Wang}},
  \citenamefont {{Weinberg}}, \citenamefont {{White}}, \citenamefont
  {{Wood-Vasey}}, \citenamefont {{Y{\`e}che}}, \citenamefont {{Zehavi}},
  \citenamefont {{Zhai}},\ and\ \citenamefont {{Zhao}}}]{Alametal:17}%
  \BibitemOpen
  \bibfield  {author} {\bibinfo {author} {\bibfnamefont {S.}~\bibnamefont
  {{Alam}}}, \bibinfo {author} {\bibfnamefont {M.}~\bibnamefont {{Ata}}},
  \bibinfo {author} {\bibfnamefont {S.}~\bibnamefont {{Bailey}}}, \bibinfo
  {author} {\bibfnamefont {F.}~\bibnamefont {{Beutler}}}, \bibinfo {author}
  {\bibfnamefont {D.}~\bibnamefont {{Bizyaev}}}, \bibinfo {author}
  {\bibfnamefont {J.~A.}\ \bibnamefont {{Blazek}}}, \bibinfo {author}
  {\bibfnamefont {A.~S.}\ \bibnamefont {{Bolton}}}, \bibinfo {author}
  {\bibfnamefont {J.~R.}\ \bibnamefont {{Brownstein}}}, \bibinfo {author}
  {\bibfnamefont {A.}~\bibnamefont {{Burden}}}, \bibinfo {author}
  {\bibfnamefont {C.-H.}\ \bibnamefont {{Chuang}}}, \bibinfo {author}
  {\bibfnamefont {J.}~\bibnamefont {{Comparat}}}, \bibinfo {author}
  {\bibfnamefont {A.~J.}\ \bibnamefont {{Cuesta}}}, \bibinfo {author}
  {\bibfnamefont {K.~S.}\ \bibnamefont {{Dawson}}}, \bibinfo {author}
  {\bibfnamefont {D.~J.}\ \bibnamefont {{Eisenstein}}}, \bibinfo {author}
  {\bibfnamefont {S.}~\bibnamefont {{Escoffier}}}, \bibinfo {author}
  {\bibfnamefont {H.}~\bibnamefont {{Gil-Mar{\'{\i}}n}}}, \bibinfo {author}
  {\bibfnamefont {J.~N.}\ \bibnamefont {{Grieb}}}, \bibinfo {author}
  {\bibfnamefont {N.}~\bibnamefont {{Hand}}}, \bibinfo {author} {\bibfnamefont
  {S.}~\bibnamefont {{Ho}}}, \bibinfo {author} {\bibfnamefont {K.}~\bibnamefont
  {{Kinemuchi}}}, \bibinfo {author} {\bibfnamefont {D.}~\bibnamefont
  {{Kirkby}}}, \bibinfo {author} {\bibfnamefont {F.}~\bibnamefont {{Kitaura}}},
  \bibinfo {author} {\bibfnamefont {E.}~\bibnamefont {{Malanushenko}}},
  \bibinfo {author} {\bibfnamefont {V.}~\bibnamefont {{Malanushenko}}},
  \bibinfo {author} {\bibfnamefont {C.}~\bibnamefont {{Maraston}}}, \bibinfo
  {author} {\bibfnamefont {C.~K.}\ \bibnamefont {{McBride}}}, \bibinfo {author}
  {\bibfnamefont {R.~C.}\ \bibnamefont {{Nichol}}}, \bibinfo {author}
  {\bibfnamefont {M.~D.}\ \bibnamefont {{Olmstead}}}, \bibinfo {author}
  {\bibfnamefont {D.}~\bibnamefont {{Oravetz}}}, \bibinfo {author}
  {\bibfnamefont {N.}~\bibnamefont {{Padmanabhan}}}, \bibinfo {author}
  {\bibfnamefont {N.}~\bibnamefont {{Palanque-Delabrouille}}}, \bibinfo
  {author} {\bibfnamefont {K.}~\bibnamefont {{Pan}}}, \bibinfo {author}
  {\bibfnamefont {M.}~\bibnamefont {{Pellejero-Ibanez}}}, \bibinfo {author}
  {\bibfnamefont {W.~J.}\ \bibnamefont {{Percival}}}, \bibinfo {author}
  {\bibfnamefont {P.}~\bibnamefont {{Petitjean}}}, \bibinfo {author}
  {\bibfnamefont {F.}~\bibnamefont {{Prada}}}, \bibinfo {author} {\bibfnamefont
  {A.~M.}\ \bibnamefont {{Price-Whelan}}}, \bibinfo {author} {\bibfnamefont
  {B.~A.}\ \bibnamefont {{Reid}}}, \bibinfo {author} {\bibfnamefont {S.~A.}\
  \bibnamefont {{Rodr{\'{\i}}guez-Torres}}}, \bibinfo {author} {\bibfnamefont
  {N.~A.}\ \bibnamefont {{Roe}}}, \bibinfo {author} {\bibfnamefont {A.~J.}\
  \bibnamefont {{Ross}}}, \bibinfo {author} {\bibfnamefont {N.~P.}\
  \bibnamefont {{Ross}}}, \bibinfo {author} {\bibfnamefont {G.}~\bibnamefont
  {{Rossi}}}, \bibinfo {author} {\bibfnamefont {J.~A.}\ \bibnamefont
  {{Rubi{\~n}o-Mart{\'{\i}}n}}}, \bibinfo {author} {\bibfnamefont
  {S.}~\bibnamefont {{Saito}}}, \bibinfo {author} {\bibfnamefont
  {S.}~\bibnamefont {{Salazar-Albornoz}}}, \bibinfo {author} {\bibfnamefont
  {L.}~\bibnamefont {{Samushia}}}, \bibinfo {author} {\bibfnamefont {A.~G.}\
  \bibnamefont {{S{\'a}nchez}}}, \bibinfo {author} {\bibfnamefont
  {S.}~\bibnamefont {{Satpathy}}}, \bibinfo {author} {\bibfnamefont {D.~J.}\
  \bibnamefont {{Schlegel}}}, \bibinfo {author} {\bibfnamefont {D.~P.}\
  \bibnamefont {{Schneider}}}, \bibinfo {author} {\bibfnamefont {C.~G.}\
  \bibnamefont {{Sc{\'o}ccola}}}, \bibinfo {author} {\bibfnamefont {H.-J.}\
  \bibnamefont {{Seo}}}, \bibinfo {author} {\bibfnamefont {E.~S.}\ \bibnamefont
  {{Sheldon}}}, \bibinfo {author} {\bibfnamefont {A.}~\bibnamefont
  {{Simmons}}}, \bibinfo {author} {\bibfnamefont {A.}~\bibnamefont {{Slosar}}},
  \bibinfo {author} {\bibfnamefont {M.~A.}\ \bibnamefont {{Strauss}}}, \bibinfo
  {author} {\bibfnamefont {M.~E.~C.}\ \bibnamefont {{Swanson}}}, \bibinfo
  {author} {\bibfnamefont {D.}~\bibnamefont {{Thomas}}}, \bibinfo {author}
  {\bibfnamefont {J.~L.}\ \bibnamefont {{Tinker}}}, \bibinfo {author}
  {\bibfnamefont {R.}~\bibnamefont {{Tojeiro}}}, \bibinfo {author}
  {\bibfnamefont {M.~V.}\ \bibnamefont {{Maga{\~n}a}}}, \bibinfo {author}
  {\bibfnamefont {J.~A.}\ \bibnamefont {{Vazquez}}}, \bibinfo {author}
  {\bibfnamefont {L.}~\bibnamefont {{Verde}}}, \bibinfo {author} {\bibfnamefont
  {D.~A.}\ \bibnamefont {{Wake}}}, \bibinfo {author} {\bibfnamefont
  {Y.}~\bibnamefont {{Wang}}}, \bibinfo {author} {\bibfnamefont {D.~H.}\
  \bibnamefont {{Weinberg}}}, \bibinfo {author} {\bibfnamefont
  {M.}~\bibnamefont {{White}}}, \bibinfo {author} {\bibfnamefont {W.~M.}\
  \bibnamefont {{Wood-Vasey}}}, \bibinfo {author} {\bibfnamefont
  {C.}~\bibnamefont {{Y{\`e}che}}}, \bibinfo {author} {\bibfnamefont
  {I.}~\bibnamefont {{Zehavi}}}, \bibinfo {author} {\bibfnamefont
  {Z.}~\bibnamefont {{Zhai}}}, \ and\ \bibinfo {author} {\bibfnamefont {G.-B.}\
  \bibnamefont {{Zhao}}},\ }\href {\doibase 10.1093/mnras/stx721} {\bibfield
  {journal} {\bibinfo  {journal} {\mnras}\ }\textbf {\bibinfo {volume} {470}},\
  \bibinfo {pages} {2617} (\bibinfo {year} {2017})},\ \Eprint
  {http://arxiv.org/abs/1607.03155} {arXiv:1607.03155} \BibitemShut {NoStop}%
\bibitem [{\citenamefont {{Dodelson}}(2003)}]{DodelsonBook}%
  \BibitemOpen
  \bibfield  {author} {\bibinfo {author} {\bibfnamefont {S.}~\bibnamefont
  {{Dodelson}}},\ }\href@noop {} {\emph {\bibinfo {title} {Modern cosmology /
  Scott Dodelson.~Amsterdam (Netherlands): Academic Press.~ISBN 0-12-219141-2,
  2003, XIII + 440 p.}}}\ (\bibinfo {year} {2003})\BibitemShut {NoStop}%
\bibitem [{\citenamefont {{Planck Collaboration}}\ \emph
  {et~al.}(2016)\citenamefont {{Planck Collaboration}}, \citenamefont {{Ade}},
  \citenamefont {{Aghanim}}, \citenamefont {{Arnaud}}, \citenamefont
  {{Ashdown}}, \citenamefont {{Aumont}}, \citenamefont {{Baccigalupi}},
  \citenamefont {{Banday}}, \citenamefont {{Barreiro}}, \citenamefont
  {{Bartlett}},\ and\ \citenamefont {et~al.}}]{PlanckCosmology:16}%
  \BibitemOpen
  \bibfield  {author} {\bibinfo {author} {\bibnamefont {{Planck
  Collaboration}}}, \bibinfo {author} {\bibfnamefont {P.~A.~R.}\ \bibnamefont
  {{Ade}}}, \bibinfo {author} {\bibfnamefont {N.}~\bibnamefont {{Aghanim}}},
  \bibinfo {author} {\bibfnamefont {M.}~\bibnamefont {{Arnaud}}}, \bibinfo
  {author} {\bibfnamefont {M.}~\bibnamefont {{Ashdown}}}, \bibinfo {author}
  {\bibfnamefont {J.}~\bibnamefont {{Aumont}}}, \bibinfo {author}
  {\bibfnamefont {C.}~\bibnamefont {{Baccigalupi}}}, \bibinfo {author}
  {\bibfnamefont {A.~J.}\ \bibnamefont {{Banday}}}, \bibinfo {author}
  {\bibfnamefont {R.~B.}\ \bibnamefont {{Barreiro}}}, \bibinfo {author}
  {\bibfnamefont {J.~G.}\ \bibnamefont {{Bartlett}}}, \ and\ \bibinfo {author}
  {\bibnamefont {et~al.}},\ }\href {\doibase 10.1051/0004-6361/201525830}
  {\bibfield  {journal} {\bibinfo  {journal} {\aap}\ }\textbf {\bibinfo
  {volume} {594}},\ \bibinfo {eid} {A13} (\bibinfo {year} {2016})},\ \Eprint
  {http://arxiv.org/abs/1502.01589} {arXiv:1502.01589} \BibitemShut {NoStop}%
\bibitem [{\citenamefont {{Bernardeau}}\ \emph {et~al.}(2002)\citenamefont
  {{Bernardeau}}, \citenamefont {{Colombi}}, \citenamefont {{Gazta{\~n}aga}},\
  and\ \citenamefont {{Scoccimarro}}}]{Bernardeauetal:02}%
  \BibitemOpen
  \bibfield  {author} {\bibinfo {author} {\bibfnamefont {F.}~\bibnamefont
  {{Bernardeau}}}, \bibinfo {author} {\bibfnamefont {S.}~\bibnamefont
  {{Colombi}}}, \bibinfo {author} {\bibfnamefont {E.}~\bibnamefont
  {{Gazta{\~n}aga}}}, \ and\ \bibinfo {author} {\bibfnamefont {R.}~\bibnamefont
  {{Scoccimarro}}},\ }\href {\doibase 10.1016/S0370-1573(02)00135-7} {\bibfield
   {journal} {\bibinfo  {journal} {\physrep}\ }\textbf {\bibinfo {volume}
  {367}},\ \bibinfo {pages} {1} (\bibinfo {year} {2002})},\ \Eprint
  {http://arxiv.org/abs/arXiv:astro-ph/0112551} {arXiv:astro-ph/0112551}
  \BibitemShut {NoStop}%
\bibitem [{\citenamefont {{Desjacques}}\ \emph {et~al.}(2016)\citenamefont
  {{Desjacques}}, \citenamefont {{Jeong}},\ and\ \citenamefont
  {{Schmidt}}}]{Desjacquesetal:16}%
  \BibitemOpen
  \bibfield  {author} {\bibinfo {author} {\bibfnamefont {V.}~\bibnamefont
  {{Desjacques}}}, \bibinfo {author} {\bibfnamefont {D.}~\bibnamefont
  {{Jeong}}}, \ and\ \bibinfo {author} {\bibfnamefont {F.}~\bibnamefont
  {{Schmidt}}},\ }\href@noop {} {\bibfield  {journal} {\bibinfo  {journal}
  {ArXiv e-prints}\ } (\bibinfo {year} {2016})},\ \Eprint
  {http://arxiv.org/abs/1611.09787} {arXiv:1611.09787} \BibitemShut {NoStop}%
\bibitem [{\citenamefont {{Takada}}\ and\ \citenamefont
  {{Jain}}(2003)}]{TakadaJain:03a}%
  \BibitemOpen
  \bibfield  {author} {\bibinfo {author} {\bibfnamefont {M.}~\bibnamefont
  {{Takada}}}\ and\ \bibinfo {author} {\bibfnamefont {B.}~\bibnamefont
  {{Jain}}},\ }\href {\doibase 10.1046/j.1365-8711.2003.06321.x} {\bibfield
  {journal} {\bibinfo  {journal} {\mnras}\ }\textbf {\bibinfo {volume} {340}},\
  \bibinfo {pages} {580} (\bibinfo {year} {2003})},\ \Eprint
  {http://arxiv.org/abs/arXiv:astro-ph/0209167} {arXiv:astro-ph/0209167}
  \BibitemShut {NoStop}%
\bibitem [{\citenamefont {{Takada}}\ and\ \citenamefont
  {{Hu}}(2013)}]{TakadaHu:13}%
  \BibitemOpen
  \bibfield  {author} {\bibinfo {author} {\bibfnamefont {M.}~\bibnamefont
  {{Takada}}}\ and\ \bibinfo {author} {\bibfnamefont {W.}~\bibnamefont
  {{Hu}}},\ }\href {\doibase 10.1103/PhysRevD.87.123504} {\bibfield  {journal}
  {\bibinfo  {journal} {\prd}\ }\textbf {\bibinfo {volume} {87}},\ \bibinfo
  {eid} {123504} (\bibinfo {year} {2013})},\ \Eprint
  {http://arxiv.org/abs/1302.6994} {arXiv:1302.6994 [astro-ph.CO]} \BibitemShut
  {NoStop}%
\bibitem [{\citenamefont {{Hamilton}}\ \emph {et~al.}(2006)\citenamefont
  {{Hamilton}}, \citenamefont {{Rimes}},\ and\ \citenamefont
  {{Scoccimarro}}}]{Hamiltonetal:06}%
  \BibitemOpen
  \bibfield  {author} {\bibinfo {author} {\bibfnamefont {A.~J.~S.}\
  \bibnamefont {{Hamilton}}}, \bibinfo {author} {\bibfnamefont {C.~D.}\
  \bibnamefont {{Rimes}}}, \ and\ \bibinfo {author} {\bibfnamefont
  {R.}~\bibnamefont {{Scoccimarro}}},\ }\href {\doibase
  10.1111/j.1365-2966.2006.10709.x} {\bibfield  {journal} {\bibinfo  {journal}
  {\mnras}\ }\textbf {\bibinfo {volume} {371}},\ \bibinfo {pages} {1188}
  (\bibinfo {year} {2006})},\ \Eprint
  {http://arxiv.org/abs/arXiv:astro-ph/0511416} {arXiv:astro-ph/0511416}
  \BibitemShut {NoStop}%
\bibitem [{\citenamefont {{Sefusatti}}\ \emph {et~al.}(2006)\citenamefont
  {{Sefusatti}}, \citenamefont {{Crocce}}, \citenamefont {{Pueblas}},\ and\
  \citenamefont {{Scoccimarro}}}]{Sefusattietal:06}%
  \BibitemOpen
  \bibfield  {author} {\bibinfo {author} {\bibfnamefont {E.}~\bibnamefont
  {{Sefusatti}}}, \bibinfo {author} {\bibfnamefont {M.}~\bibnamefont
  {{Crocce}}}, \bibinfo {author} {\bibfnamefont {S.}~\bibnamefont {{Pueblas}}},
  \ and\ \bibinfo {author} {\bibfnamefont {R.}~\bibnamefont {{Scoccimarro}}},\
  }\href {\doibase 10.1103/PhysRevD.74.023522} {\bibfield  {journal} {\bibinfo
  {journal} {\prd}\ }\textbf {\bibinfo {volume} {74}},\ \bibinfo {eid} {023522}
  (\bibinfo {year} {2006})},\ \Eprint
  {http://arxiv.org/abs/arXiv:astro-ph/0604505} {arXiv:astro-ph/0604505}
  \BibitemShut {NoStop}%
\bibitem [{\citenamefont {{Takada}}\ and\ \citenamefont
  {{Bridle}}(2007)}]{TakadaBridle:07}%
  \BibitemOpen
  \bibfield  {author} {\bibinfo {author} {\bibfnamefont {M.}~\bibnamefont
  {{Takada}}}\ and\ \bibinfo {author} {\bibfnamefont {S.}~\bibnamefont
  {{Bridle}}},\ }\href {\doibase 10.1088/1367-2630/9/12/446} {\bibfield
  {journal} {\bibinfo  {journal} {New Journal of Physics}\ }\textbf {\bibinfo
  {volume} {9}},\ \bibinfo {pages} {446} (\bibinfo {year} {2007})},\ \Eprint
  {http://arxiv.org/abs/arXiv:0705.0163} {arXiv:arXiv:0705.0163} \BibitemShut
  {NoStop}%
\bibitem [{\citenamefont {{Takada}}\ and\ \citenamefont
  {{Jain}}(2009)}]{TakadaJain:09}%
  \BibitemOpen
  \bibfield  {author} {\bibinfo {author} {\bibfnamefont {M.}~\bibnamefont
  {{Takada}}}\ and\ \bibinfo {author} {\bibfnamefont {B.}~\bibnamefont
  {{Jain}}},\ }\href {\doibase 10.1111/j.1365-2966.2009.14504.x} {\bibfield
  {journal} {\bibinfo  {journal} {\mnras}\ }\textbf {\bibinfo {volume} {395}},\
  \bibinfo {pages} {2065} (\bibinfo {year} {2009})},\ \Eprint
  {http://arxiv.org/abs/0810.4170} {arXiv:0810.4170} \BibitemShut {NoStop}%
\bibitem [{\citenamefont {{Sato}}\ \emph {et~al.}(2009)\citenamefont {{Sato}},
  \citenamefont {{Hamana}}, \citenamefont {{Takahashi}}, \citenamefont
  {{Takada}}, \citenamefont {{Yoshida}}, \citenamefont {{Matsubara}},\ and\
  \citenamefont {{Sugiyama}}}]{Satoetal:09}%
  \BibitemOpen
  \bibfield  {author} {\bibinfo {author} {\bibfnamefont {M.}~\bibnamefont
  {{Sato}}}, \bibinfo {author} {\bibfnamefont {T.}~\bibnamefont {{Hamana}}},
  \bibinfo {author} {\bibfnamefont {R.}~\bibnamefont {{Takahashi}}}, \bibinfo
  {author} {\bibfnamefont {M.}~\bibnamefont {{Takada}}}, \bibinfo {author}
  {\bibfnamefont {N.}~\bibnamefont {{Yoshida}}}, \bibinfo {author}
  {\bibfnamefont {T.}~\bibnamefont {{Matsubara}}}, \ and\ \bibinfo {author}
  {\bibfnamefont {N.}~\bibnamefont {{Sugiyama}}},\ }\href {\doibase
  10.1088/0004-637X/701/2/945} {\bibfield  {journal} {\bibinfo  {journal}
  {\apj}\ }\textbf {\bibinfo {volume} {701}},\ \bibinfo {pages} {945} (\bibinfo
  {year} {2009})},\ \Eprint {http://arxiv.org/abs/0906.2237} {arXiv:0906.2237
  [astro-ph.CO]} \BibitemShut {NoStop}%
\bibitem [{\citenamefont {{Baldauf}}\ \emph {et~al.}(2011)\citenamefont
  {{Baldauf}}, \citenamefont {{Seljak}}, \citenamefont {{Senatore}},\ and\
  \citenamefont {{Zaldarriaga}}}]{Baldaufetal:11}%
  \BibitemOpen
  \bibfield  {author} {\bibinfo {author} {\bibfnamefont {T.}~\bibnamefont
  {{Baldauf}}}, \bibinfo {author} {\bibfnamefont {U.}~\bibnamefont {{Seljak}}},
  \bibinfo {author} {\bibfnamefont {L.}~\bibnamefont {{Senatore}}}, \ and\
  \bibinfo {author} {\bibfnamefont {M.}~\bibnamefont {{Zaldarriaga}}},\ }\href
  {\doibase 10.1088/1475-7516/2011/10/031} {\bibfield  {journal} {\bibinfo
  {journal} {\jcap}\ }\textbf {\bibinfo {volume} {10}},\ \bibinfo {eid} {031}
  (\bibinfo {year} {2011})},\ \Eprint {http://arxiv.org/abs/1106.5507}
  {arXiv:1106.5507} \BibitemShut {NoStop}%
\bibitem [{\citenamefont {{Sherwin}}\ and\ \citenamefont
  {{Zaldarriaga}}(2012)}]{SherwinZaldarriaga:12}%
  \BibitemOpen
  \bibfield  {author} {\bibinfo {author} {\bibfnamefont {B.~D.}\ \bibnamefont
  {{Sherwin}}}\ and\ \bibinfo {author} {\bibfnamefont {M.}~\bibnamefont
  {{Zaldarriaga}}},\ }\href {\doibase 10.1103/PhysRevD.85.103523} {\bibfield
  {journal} {\bibinfo  {journal} {\prd}\ }\textbf {\bibinfo {volume} {85}},\
  \bibinfo {eid} {103523} (\bibinfo {year} {2012})},\ \Eprint
  {http://arxiv.org/abs/1202.3998} {arXiv:1202.3998 [astro-ph.CO]} \BibitemShut
  {NoStop}%
\bibitem [{\citenamefont {{de Putter}}\ \emph {et~al.}(2012)\citenamefont {{de
  Putter}}, \citenamefont {{Wagner}}, \citenamefont {{Mena}}, \citenamefont
  {{Verde}},\ and\ \citenamefont {{Percival}}}]{dePutteretal:12}%
  \BibitemOpen
  \bibfield  {author} {\bibinfo {author} {\bibfnamefont {R.}~\bibnamefont {{de
  Putter}}}, \bibinfo {author} {\bibfnamefont {C.}~\bibnamefont {{Wagner}}},
  \bibinfo {author} {\bibfnamefont {O.}~\bibnamefont {{Mena}}}, \bibinfo
  {author} {\bibfnamefont {L.}~\bibnamefont {{Verde}}}, \ and\ \bibinfo
  {author} {\bibfnamefont {W.~J.}\ \bibnamefont {{Percival}}},\ }\href
  {\doibase 10.1088/1475-7516/2012/04/019} {\bibfield  {journal} {\bibinfo
  {journal} {\jcap}\ }\textbf {\bibinfo {volume} {4}},\ \bibinfo {eid} {019}
  (\bibinfo {year} {2012})},\ \Eprint {http://arxiv.org/abs/1111.6596}
  {arXiv:1111.6596} \BibitemShut {NoStop}%
\bibitem [{\citenamefont {{Takada}}\ and\ \citenamefont
  {{Spergel}}(2013)}]{TakadaSpergel:13}%
  \BibitemOpen
  \bibfield  {author} {\bibinfo {author} {\bibfnamefont {M.}~\bibnamefont
  {{Takada}}}\ and\ \bibinfo {author} {\bibfnamefont {D.~N.}\ \bibnamefont
  {{Spergel}}},\ }\href@noop {} {\bibfield  {journal} {\bibinfo  {journal}
  {ArXiv e-prints}\ } (\bibinfo {year} {2013})},\ \Eprint
  {http://arxiv.org/abs/1307.4399} {arXiv:1307.4399 [astro-ph.CO]} \BibitemShut
  {NoStop}%
\bibitem [{\citenamefont {{Schaan}}\ \emph {et~al.}(2014)\citenamefont
  {{Schaan}}, \citenamefont {{Takada}},\ and\ \citenamefont
  {{Spergel}}}]{Schaanetal:14}%
  \BibitemOpen
  \bibfield  {author} {\bibinfo {author} {\bibfnamefont {E.}~\bibnamefont
  {{Schaan}}}, \bibinfo {author} {\bibfnamefont {M.}~\bibnamefont {{Takada}}},
  \ and\ \bibinfo {author} {\bibfnamefont {D.~N.}\ \bibnamefont {{Spergel}}},\
  }\href@noop {} {\bibfield  {journal} {\bibinfo  {journal} {ArXiv e-prints}\ }
  (\bibinfo {year} {2014})},\ \Eprint {http://arxiv.org/abs/1406.3330}
  {arXiv:1406.3330} \BibitemShut {NoStop}%
\bibitem [{\citenamefont {{Li}}\ \emph
  {et~al.}(2014{\natexlab{a}})\citenamefont {{Li}}, \citenamefont {{Hu}},\ and\
  \citenamefont {{Takada}}}]{Lietal:14a}%
  \BibitemOpen
  \bibfield  {author} {\bibinfo {author} {\bibfnamefont {Y.}~\bibnamefont
  {{Li}}}, \bibinfo {author} {\bibfnamefont {W.}~\bibnamefont {{Hu}}}, \ and\
  \bibinfo {author} {\bibfnamefont {M.}~\bibnamefont {{Takada}}},\ }\href
  {\doibase 10.1103/PhysRevD.89.083519} {\bibfield  {journal} {\bibinfo
  {journal} {\prd}\ }\textbf {\bibinfo {volume} {89}},\ \bibinfo {eid} {083519}
  (\bibinfo {year} {2014}{\natexlab{a}})},\ \Eprint
  {http://arxiv.org/abs/1401.0385} {arXiv:1401.0385} \BibitemShut {NoStop}%
\bibitem [{\citenamefont {{Li}}\ \emph
  {et~al.}(2014{\natexlab{b}})\citenamefont {{Li}}, \citenamefont {{Hu}},\ and\
  \citenamefont {{Takada}}}]{Lietal:14b}%
  \BibitemOpen
  \bibfield  {author} {\bibinfo {author} {\bibfnamefont {Y.}~\bibnamefont
  {{Li}}}, \bibinfo {author} {\bibfnamefont {W.}~\bibnamefont {{Hu}}}, \ and\
  \bibinfo {author} {\bibfnamefont {M.}~\bibnamefont {{Takada}}},\ }\href
  {\doibase 10.1103/PhysRevD.90.103530} {\bibfield  {journal} {\bibinfo
  {journal} {\prd}\ }\textbf {\bibinfo {volume} {90}},\ \bibinfo {eid} {103530}
  (\bibinfo {year} {2014}{\natexlab{b}})},\ \Eprint
  {http://arxiv.org/abs/1408.1081} {arXiv:1408.1081} \BibitemShut {NoStop}%
\bibitem [{\citenamefont {{Takahashi}}\ \emph {et~al.}(2014)\citenamefont
  {{Takahashi}}, \citenamefont {{Soma}}, \citenamefont {{Takada}},\ and\
  \citenamefont {{Kayo}}}]{Takahashietal:14}%
  \BibitemOpen
  \bibfield  {author} {\bibinfo {author} {\bibfnamefont {R.}~\bibnamefont
  {{Takahashi}}}, \bibinfo {author} {\bibfnamefont {S.}~\bibnamefont {{Soma}}},
  \bibinfo {author} {\bibfnamefont {M.}~\bibnamefont {{Takada}}}, \ and\
  \bibinfo {author} {\bibfnamefont {I.}~\bibnamefont {{Kayo}}},\ }\href
  {\doibase 10.1093/mnras/stu1693} {\bibfield  {journal} {\bibinfo  {journal}
  {\mnras}\ }\textbf {\bibinfo {volume} {444}},\ \bibinfo {pages} {3473}
  (\bibinfo {year} {2014})},\ \Eprint {http://arxiv.org/abs/1405.2666}
  {arXiv:1405.2666} \BibitemShut {NoStop}%
\bibitem [{\citenamefont {{Manzotti}}\ \emph {et~al.}(2014)\citenamefont
  {{Manzotti}}, \citenamefont {{Hu}},\ and\ \citenamefont
  {{Benoit-L{\'e}vy}}}]{Manzottietal:14}%
  \BibitemOpen
  \bibfield  {author} {\bibinfo {author} {\bibfnamefont {A.}~\bibnamefont
  {{Manzotti}}}, \bibinfo {author} {\bibfnamefont {W.}~\bibnamefont {{Hu}}}, \
  and\ \bibinfo {author} {\bibfnamefont {A.}~\bibnamefont
  {{Benoit-L{\'e}vy}}},\ }\href {\doibase 10.1103/PhysRevD.90.023003}
  {\bibfield  {journal} {\bibinfo  {journal} {\prd}\ }\textbf {\bibinfo
  {volume} {90}},\ \bibinfo {eid} {023003} (\bibinfo {year} {2014})},\ \Eprint
  {http://arxiv.org/abs/1401.7992} {arXiv:1401.7992} \BibitemShut {NoStop}%
\bibitem [{\citenamefont {{Carron}}\ and\ \citenamefont
  {{Szapudi}}(2015)}]{CarronSzapudi:15}%
  \BibitemOpen
  \bibfield  {author} {\bibinfo {author} {\bibfnamefont {J.}~\bibnamefont
  {{Carron}}}\ and\ \bibinfo {author} {\bibfnamefont {I.}~\bibnamefont
  {{Szapudi}}},\ }\href {\doibase 10.1093/mnras/stu2501} {\bibfield  {journal}
  {\bibinfo  {journal} {\mnras}\ }\textbf {\bibinfo {volume} {447}},\ \bibinfo
  {pages} {671} (\bibinfo {year} {2015})},\ \Eprint
  {http://arxiv.org/abs/1408.1744} {arXiv:1408.1744} \BibitemShut {NoStop}%
\bibitem [{\citenamefont {{Dai}}\ \emph {et~al.}(2015)\citenamefont {{Dai}},
  \citenamefont {{Pajer}},\ and\ \citenamefont {{Schmidt}}}]{Daietal:15}%
  \BibitemOpen
  \bibfield  {author} {\bibinfo {author} {\bibfnamefont {L.}~\bibnamefont
  {{Dai}}}, \bibinfo {author} {\bibfnamefont {E.}~\bibnamefont {{Pajer}}}, \
  and\ \bibinfo {author} {\bibfnamefont {F.}~\bibnamefont {{Schmidt}}},\ }\href
  {\doibase 10.1088/1475-7516/2015/10/059} {\bibfield  {journal} {\bibinfo
  {journal} {\jcap}\ }\textbf {\bibinfo {volume} {10}},\ \bibinfo {eid} {059}
  (\bibinfo {year} {2015})},\ \Eprint {http://arxiv.org/abs/1504.00351}
  {arXiv:1504.00351} \BibitemShut {NoStop}%
\bibitem [{\citenamefont {{Shirasaki}}\ \emph {et~al.}(2017)\citenamefont
  {{Shirasaki}}, \citenamefont {{Takada}}, \citenamefont {{Miyatake}},
  \citenamefont {{Takahashi}}, \citenamefont {{Hamana}}, \citenamefont
  {{Nishimichi}},\ and\ \citenamefont {{Murata}}}]{Shirasakietal:17}%
  \BibitemOpen
  \bibfield  {author} {\bibinfo {author} {\bibfnamefont {M.}~\bibnamefont
  {{Shirasaki}}}, \bibinfo {author} {\bibfnamefont {M.}~\bibnamefont
  {{Takada}}}, \bibinfo {author} {\bibfnamefont {H.}~\bibnamefont
  {{Miyatake}}}, \bibinfo {author} {\bibfnamefont {R.}~\bibnamefont
  {{Takahashi}}}, \bibinfo {author} {\bibfnamefont {T.}~\bibnamefont
  {{Hamana}}}, \bibinfo {author} {\bibfnamefont {T.}~\bibnamefont
  {{Nishimichi}}}, \ and\ \bibinfo {author} {\bibfnamefont {R.}~\bibnamefont
  {{Murata}}},\ }\href {\doibase 10.1093/mnras/stx1477} {\bibfield  {journal}
  {\bibinfo  {journal} {\mnras}\ }\textbf {\bibinfo {volume} {470}},\ \bibinfo
  {pages} {3476} (\bibinfo {year} {2017})},\ \Eprint
  {http://arxiv.org/abs/1607.08679} {arXiv:1607.08679} \BibitemShut {NoStop}%
\bibitem [{\citenamefont {{Li}}\ \emph {et~al.}(2016)\citenamefont {{Li}},
  \citenamefont {{Hu}},\ and\ \citenamefont {{Takada}}}]{Lietal:16}%
  \BibitemOpen
  \bibfield  {author} {\bibinfo {author} {\bibfnamefont {Y.}~\bibnamefont
  {{Li}}}, \bibinfo {author} {\bibfnamefont {W.}~\bibnamefont {{Hu}}}, \ and\
  \bibinfo {author} {\bibfnamefont {M.}~\bibnamefont {{Takada}}},\ }\href
  {\doibase 10.1103/PhysRevD.93.063507} {\bibfield  {journal} {\bibinfo
  {journal} {\prd}\ }\textbf {\bibinfo {volume} {93}},\ \bibinfo {eid} {063507}
  (\bibinfo {year} {2016})},\ \Eprint {http://arxiv.org/abs/1511.01454}
  {arXiv:1511.01454} \BibitemShut {NoStop}%
\bibitem [{\citenamefont {{Ip}}\ and\ \citenamefont
  {{Schmidt}}(2017)}]{IpSchmidt:17}%
  \BibitemOpen
  \bibfield  {author} {\bibinfo {author} {\bibfnamefont {H.~Y.}\ \bibnamefont
  {{Ip}}}\ and\ \bibinfo {author} {\bibfnamefont {F.}~\bibnamefont
  {{Schmidt}}},\ }\href {\doibase 10.1088/1475-7516/2017/02/025} {\bibfield
  {journal} {\bibinfo  {journal} {\jcap}\ }\textbf {\bibinfo {volume} {2}},\
  \bibinfo {eid} {025} (\bibinfo {year} {2017})},\ \Eprint
  {http://arxiv.org/abs/1610.01059} {arXiv:1610.01059} \BibitemShut {NoStop}%
\bibitem [{\citenamefont {{Akitsu}}\ \emph {et~al.}(2017)\citenamefont
  {{Akitsu}}, \citenamefont {{Takada}},\ and\ \citenamefont
  {{Li}}}]{Akitsuetal:17}%
  \BibitemOpen
  \bibfield  {author} {\bibinfo {author} {\bibfnamefont {K.}~\bibnamefont
  {{Akitsu}}}, \bibinfo {author} {\bibfnamefont {M.}~\bibnamefont {{Takada}}},
  \ and\ \bibinfo {author} {\bibfnamefont {Y.}~\bibnamefont {{Li}}},\ }\href
  {\doibase 10.1103/PhysRevD.95.083522} {\bibfield  {journal} {\bibinfo
  {journal} {\prd}\ }\textbf {\bibinfo {volume} {95}},\ \bibinfo {eid} {083522}
  (\bibinfo {year} {2017})},\ \Eprint {http://arxiv.org/abs/1611.04723}
  {arXiv:1611.04723} \BibitemShut {NoStop}%
\bibitem [{\citenamefont {{Barreira}}\ and\ \citenamefont
  {{Schmidt}}(2017)}]{BarreiraSchmidt:17a}%
  \BibitemOpen
  \bibfield  {author} {\bibinfo {author} {\bibfnamefont {A.}~\bibnamefont
  {{Barreira}}}\ and\ \bibinfo {author} {\bibfnamefont {F.}~\bibnamefont
  {{Schmidt}}},\ }\href@noop {} {\bibfield  {journal} {\bibinfo  {journal}
  {ArXiv e-prints}\ } (\bibinfo {year} {2017})},\ \Eprint
  {http://arxiv.org/abs/1703.09212} {arXiv:1703.09212} \BibitemShut {NoStop}%
\bibitem [{\citenamefont {{Krause}}\ \emph {et~al.}(2017)\citenamefont
  {{Krause}}, \citenamefont {{Eifler}}, \citenamefont {{Zuntz}}, \citenamefont
  {{Friedrich}}, \citenamefont {{Troxel}}, \citenamefont {{Dodelson}},
  \citenamefont {{Blazek}}, \citenamefont {{Secco}}, \citenamefont
  {{MacCrann}}, \citenamefont {{Baxter}}, \citenamefont {{Chang}},
  \citenamefont {{Chen}}, \citenamefont {{Crocce}}, \citenamefont {{DeRose}},
  \citenamefont {{Ferte}}, \citenamefont {{Kokron}}, \citenamefont {{Lacasa}},
  \citenamefont {{Miranda}}, \citenamefont {{Omori}}, \citenamefont
  {{Porredon}}, \citenamefont {{Rosenfeld}}, \citenamefont {{Samuroff}},
  \citenamefont {{Wang}}, \citenamefont {{Wechsler}}, \citenamefont {{Abbott}},
  \citenamefont {{Abdalla}}, \citenamefont {{Allam}}, \citenamefont {{Annis}},
  \citenamefont {{Bechtol}}, \citenamefont {{Benoit-Levy}}, \citenamefont
  {{Bernstein}}, \citenamefont {{Brooks}}, \citenamefont {{Burke}},
  \citenamefont {{Capozzi}}, \citenamefont {{Carrasco Kind}}, \citenamefont
  {{Carretero}}, \citenamefont {{D'Andrea}}, \citenamefont {{da Costa}},
  \citenamefont {{Davis}}, \citenamefont {{DePoy}}, \citenamefont {{Desai}},
  \citenamefont {{Diehl}}, \citenamefont {{Dietrich}}, \citenamefont
  {{Evrard}}, \citenamefont {{Flaugher}}, \citenamefont {{Fosalba}},
  \citenamefont {{Frieman}}, \citenamefont {{Garcia-Bellido}}, \citenamefont
  {{Gaztanaga}}, \citenamefont {{Giannantonio}}, \citenamefont {{Gruen}},
  \citenamefont {{Gruendl}}, \citenamefont {{Gschwend}}, \citenamefont
  {{Gutierrez}}, \citenamefont {{Honscheid}}, \citenamefont {{James}},
  \citenamefont {{Jeltema}}, \citenamefont {{Kuehn}}, \citenamefont
  {{Kuhlmann}}, \citenamefont {{Lahav}}, \citenamefont {{Lima}}, \citenamefont
  {{Maia}}, \citenamefont {{March}}, \citenamefont {{Marshall}}, \citenamefont
  {{Martini}}, \citenamefont {{Menanteau}}, \citenamefont {{Miquel}},
  \citenamefont {{Nichol}}, \citenamefont {{Plazas}}, \citenamefont {{Romer}},
  \citenamefont {{Rykoff}}, \citenamefont {{Sanchez}}, \citenamefont
  {{Scarpine}}, \citenamefont {{Schindler}}, \citenamefont {{Schubnell}},
  \citenamefont {{Sevilla-Noarbe}}, \citenamefont {{Smith}}, \citenamefont
  {{Soares-Santos}}, \citenamefont {{Sobreira}}, \citenamefont {{Suchyta}},
  \citenamefont {{Swanson}}, \citenamefont {{Tarle}}, \citenamefont {{Tucker}},
  \citenamefont {{Vikram}}, \citenamefont {{Walker}},\ and\ \citenamefont
  {{Weller}}}]{Kuraseetal:17}%
  \BibitemOpen
  \bibfield  {author} {\bibinfo {author} {\bibfnamefont {E.}~\bibnamefont
  {{Krause}}}, \bibinfo {author} {\bibfnamefont {T.~F.}\ \bibnamefont
  {{Eifler}}}, \bibinfo {author} {\bibfnamefont {J.}~\bibnamefont {{Zuntz}}},
  \bibinfo {author} {\bibfnamefont {O.}~\bibnamefont {{Friedrich}}}, \bibinfo
  {author} {\bibfnamefont {M.~A.}\ \bibnamefont {{Troxel}}}, \bibinfo {author}
  {\bibfnamefont {S.}~\bibnamefont {{Dodelson}}}, \bibinfo {author}
  {\bibfnamefont {J.}~\bibnamefont {{Blazek}}}, \bibinfo {author}
  {\bibfnamefont {L.~F.}\ \bibnamefont {{Secco}}}, \bibinfo {author}
  {\bibfnamefont {N.}~\bibnamefont {{MacCrann}}}, \bibinfo {author}
  {\bibfnamefont {E.}~\bibnamefont {{Baxter}}}, \bibinfo {author}
  {\bibfnamefont {C.}~\bibnamefont {{Chang}}}, \bibinfo {author} {\bibfnamefont
  {N.}~\bibnamefont {{Chen}}}, \bibinfo {author} {\bibfnamefont
  {M.}~\bibnamefont {{Crocce}}}, \bibinfo {author} {\bibfnamefont
  {J.}~\bibnamefont {{DeRose}}}, \bibinfo {author} {\bibfnamefont
  {A.}~\bibnamefont {{Ferte}}}, \bibinfo {author} {\bibfnamefont
  {N.}~\bibnamefont {{Kokron}}}, \bibinfo {author} {\bibfnamefont
  {F.}~\bibnamefont {{Lacasa}}}, \bibinfo {author} {\bibfnamefont
  {V.}~\bibnamefont {{Miranda}}}, \bibinfo {author} {\bibfnamefont
  {Y.}~\bibnamefont {{Omori}}}, \bibinfo {author} {\bibfnamefont
  {A.}~\bibnamefont {{Porredon}}}, \bibinfo {author} {\bibfnamefont
  {R.}~\bibnamefont {{Rosenfeld}}}, \bibinfo {author} {\bibfnamefont
  {S.}~\bibnamefont {{Samuroff}}}, \bibinfo {author} {\bibfnamefont
  {M.}~\bibnamefont {{Wang}}}, \bibinfo {author} {\bibfnamefont {R.~H.}\
  \bibnamefont {{Wechsler}}}, \bibinfo {author} {\bibfnamefont {T.~M.~C.}\
  \bibnamefont {{Abbott}}}, \bibinfo {author} {\bibfnamefont {F.~B.}\
  \bibnamefont {{Abdalla}}}, \bibinfo {author} {\bibfnamefont {S.}~\bibnamefont
  {{Allam}}}, \bibinfo {author} {\bibfnamefont {J.}~\bibnamefont {{Annis}}},
  \bibinfo {author} {\bibfnamefont {K.}~\bibnamefont {{Bechtol}}}, \bibinfo
  {author} {\bibfnamefont {A.}~\bibnamefont {{Benoit-Levy}}}, \bibinfo {author}
  {\bibfnamefont {G.~M.}\ \bibnamefont {{Bernstein}}}, \bibinfo {author}
  {\bibfnamefont {D.}~\bibnamefont {{Brooks}}}, \bibinfo {author}
  {\bibfnamefont {D.~L.}\ \bibnamefont {{Burke}}}, \bibinfo {author}
  {\bibfnamefont {D.}~\bibnamefont {{Capozzi}}}, \bibinfo {author}
  {\bibfnamefont {M.}~\bibnamefont {{Carrasco Kind}}}, \bibinfo {author}
  {\bibfnamefont {J.}~\bibnamefont {{Carretero}}}, \bibinfo {author}
  {\bibfnamefont {C.~B.}\ \bibnamefont {{D'Andrea}}}, \bibinfo {author}
  {\bibfnamefont {L.~N.}\ \bibnamefont {{da Costa}}}, \bibinfo {author}
  {\bibfnamefont {C.}~\bibnamefont {{Davis}}}, \bibinfo {author} {\bibfnamefont
  {D.~L.}\ \bibnamefont {{DePoy}}}, \bibinfo {author} {\bibfnamefont
  {S.}~\bibnamefont {{Desai}}}, \bibinfo {author} {\bibfnamefont {H.~T.}\
  \bibnamefont {{Diehl}}}, \bibinfo {author} {\bibfnamefont {J.~P.}\
  \bibnamefont {{Dietrich}}}, \bibinfo {author} {\bibfnamefont {A.~E.}\
  \bibnamefont {{Evrard}}}, \bibinfo {author} {\bibfnamefont {B.}~\bibnamefont
  {{Flaugher}}}, \bibinfo {author} {\bibfnamefont {P.}~\bibnamefont
  {{Fosalba}}}, \bibinfo {author} {\bibfnamefont {J.}~\bibnamefont
  {{Frieman}}}, \bibinfo {author} {\bibfnamefont {J.}~\bibnamefont
  {{Garcia-Bellido}}}, \bibinfo {author} {\bibfnamefont {E.}~\bibnamefont
  {{Gaztanaga}}}, \bibinfo {author} {\bibfnamefont {T.}~\bibnamefont
  {{Giannantonio}}}, \bibinfo {author} {\bibfnamefont {D.}~\bibnamefont
  {{Gruen}}}, \bibinfo {author} {\bibfnamefont {R.~A.}\ \bibnamefont
  {{Gruendl}}}, \bibinfo {author} {\bibfnamefont {J.}~\bibnamefont
  {{Gschwend}}}, \bibinfo {author} {\bibfnamefont {G.}~\bibnamefont
  {{Gutierrez}}}, \bibinfo {author} {\bibfnamefont {K.}~\bibnamefont
  {{Honscheid}}}, \bibinfo {author} {\bibfnamefont {D.~J.}\ \bibnamefont
  {{James}}}, \bibinfo {author} {\bibfnamefont {T.}~\bibnamefont {{Jeltema}}},
  \bibinfo {author} {\bibfnamefont {K.}~\bibnamefont {{Kuehn}}}, \bibinfo
  {author} {\bibfnamefont {S.}~\bibnamefont {{Kuhlmann}}}, \bibinfo {author}
  {\bibfnamefont {O.}~\bibnamefont {{Lahav}}}, \bibinfo {author} {\bibfnamefont
  {M.}~\bibnamefont {{Lima}}}, \bibinfo {author} {\bibfnamefont {M.~A.~G.}\
  \bibnamefont {{Maia}}}, \bibinfo {author} {\bibfnamefont {M.}~\bibnamefont
  {{March}}}, \bibinfo {author} {\bibfnamefont {J.~L.}\ \bibnamefont
  {{Marshall}}}, \bibinfo {author} {\bibfnamefont {P.}~\bibnamefont
  {{Martini}}}, \bibinfo {author} {\bibfnamefont {F.}~\bibnamefont
  {{Menanteau}}}, \bibinfo {author} {\bibfnamefont {R.}~\bibnamefont
  {{Miquel}}}, \bibinfo {author} {\bibfnamefont {R.~C.}\ \bibnamefont
  {{Nichol}}}, \bibinfo {author} {\bibfnamefont {A.~A.}\ \bibnamefont
  {{Plazas}}}, \bibinfo {author} {\bibfnamefont {A.~K.}\ \bibnamefont
  {{Romer}}}, \bibinfo {author} {\bibfnamefont {E.~S.}\ \bibnamefont
  {{Rykoff}}}, \bibinfo {author} {\bibfnamefont {E.}~\bibnamefont {{Sanchez}}},
  \bibinfo {author} {\bibfnamefont {V.}~\bibnamefont {{Scarpine}}}, \bibinfo
  {author} {\bibfnamefont {R.}~\bibnamefont {{Schindler}}}, \bibinfo {author}
  {\bibfnamefont {M.}~\bibnamefont {{Schubnell}}}, \bibinfo {author}
  {\bibfnamefont {I.}~\bibnamefont {{Sevilla-Noarbe}}}, \bibinfo {author}
  {\bibfnamefont {M.}~\bibnamefont {{Smith}}}, \bibinfo {author} {\bibfnamefont
  {M.}~\bibnamefont {{Soares-Santos}}}, \bibinfo {author} {\bibfnamefont
  {F.}~\bibnamefont {{Sobreira}}}, \bibinfo {author} {\bibfnamefont
  {E.}~\bibnamefont {{Suchyta}}}, \bibinfo {author} {\bibfnamefont {M.~E.~C.}\
  \bibnamefont {{Swanson}}}, \bibinfo {author} {\bibfnamefont {G.}~\bibnamefont
  {{Tarle}}}, \bibinfo {author} {\bibfnamefont {D.~L.}\ \bibnamefont
  {{Tucker}}}, \bibinfo {author} {\bibfnamefont {V.}~\bibnamefont {{Vikram}}},
  \bibinfo {author} {\bibfnamefont {A.~R.}\ \bibnamefont {{Walker}}}, \ and\
  \bibinfo {author} {\bibfnamefont {J.}~\bibnamefont {{Weller}}},\ }\href@noop
  {} {\bibfield  {journal} {\bibinfo  {journal} {ArXiv e-prints}\ } (\bibinfo
  {year} {2017})},\ \Eprint {http://arxiv.org/abs/1706.09359}
  {arXiv:1706.09359} \BibitemShut {NoStop}%
\bibitem [{\citenamefont {{Schmidt}}\ \emph {et~al.}(2014)\citenamefont
  {{Schmidt}}, \citenamefont {{Pajer}},\ and\ \citenamefont
  {{Zaldarriaga}}}]{Schmidtetal:14}%
  \BibitemOpen
  \bibfield  {author} {\bibinfo {author} {\bibfnamefont {F.}~\bibnamefont
  {{Schmidt}}}, \bibinfo {author} {\bibfnamefont {E.}~\bibnamefont {{Pajer}}},
  \ and\ \bibinfo {author} {\bibfnamefont {M.}~\bibnamefont {{Zaldarriaga}}},\
  }\href {\doibase 10.1103/PhysRevD.89.083507} {\bibfield  {journal} {\bibinfo
  {journal} {\prd}\ }\textbf {\bibinfo {volume} {89}},\ \bibinfo {eid} {083507}
  (\bibinfo {year} {2014})},\ \Eprint {http://arxiv.org/abs/1312.5616}
  {arXiv:1312.5616} \BibitemShut {NoStop}%
\bibitem [{\citenamefont {{Kaiser}}(1987)}]{Kaiser:87}%
  \BibitemOpen
  \bibfield  {author} {\bibinfo {author} {\bibfnamefont {N.}~\bibnamefont
  {{Kaiser}}},\ }\href@noop {} {\bibfield  {journal} {\bibinfo  {journal}
  {\mnras}\ }\textbf {\bibinfo {volume} {227}},\ \bibinfo {pages} {1} (\bibinfo
  {year} {1987})}\BibitemShut {NoStop}%
\bibitem [{\citenamefont {{Hamilton}}(1998)}]{Hamilton:98}%
  \BibitemOpen
  \bibfield  {author} {\bibinfo {author} {\bibfnamefont {A.~J.~S.}\
  \bibnamefont {{Hamilton}}},\ }in\ \href@noop {} {\emph {\bibinfo {booktitle}
  {The Evolving Universe}}},\ \bibinfo {series} {Astrophysics and Space Science
  Library}, Vol.\ \bibinfo {volume} {231},\ \bibinfo {editor} {edited by\
  \bibinfo {editor} {\bibnamefont {{D.~Hamilton}}}}\ (\bibinfo {year} {1998})\
  pp.\ \bibinfo {pages} {185--+}\BibitemShut {NoStop}%
\bibitem [{\citenamefont {{Alcock}}\ and\ \citenamefont
  {{Paczynski}}(1979)}]{AlcockPaczynski:79}%
  \BibitemOpen
  \bibfield  {author} {\bibinfo {author} {\bibfnamefont {C.}~\bibnamefont
  {{Alcock}}}\ and\ \bibinfo {author} {\bibfnamefont {B.}~\bibnamefont
  {{Paczynski}}},\ }\href {\doibase 10.1038/281358a0} {\bibfield  {journal}
  {\bibinfo  {journal} {\nat}\ }\textbf {\bibinfo {volume} {281}},\ \bibinfo
  {pages} {358} (\bibinfo {year} {1979})}\BibitemShut {NoStop}%
\bibitem [{\citenamefont {{Seo}}\ and\ \citenamefont
  {{Eisenstein}}(2003)}]{SeoEisenstein:03}%
  \BibitemOpen
  \bibfield  {author} {\bibinfo {author} {\bibfnamefont {H.}~\bibnamefont
  {{Seo}}}\ and\ \bibinfo {author} {\bibfnamefont {D.~J.}\ \bibnamefont
  {{Eisenstein}}},\ }\href {\doibase 10.1086/379122} {\bibfield  {journal}
  {\bibinfo  {journal} {\apj}\ }\textbf {\bibinfo {volume} {598}},\ \bibinfo
  {pages} {720} (\bibinfo {year} {2003})},\ \Eprint
  {http://arxiv.org/abs/arXiv:astro-ph/0307460} {arXiv:astro-ph/0307460}
  \BibitemShut {NoStop}%
\bibitem [{\citenamefont {{Hu}}\ and\ \citenamefont
  {{Haiman}}(2003)}]{HuHaiman:03}%
  \BibitemOpen
  \bibfield  {author} {\bibinfo {author} {\bibfnamefont {W.}~\bibnamefont
  {{Hu}}}\ and\ \bibinfo {author} {\bibfnamefont {Z.}~\bibnamefont
  {{Haiman}}},\ }\href {\doibase 10.1103/PhysRevD.68.063004} {\bibfield
  {journal} {\bibinfo  {journal} {\prd}\ }\textbf {\bibinfo {volume} {68}},\
  \bibinfo {eid} {063004} (\bibinfo {year} {2003})},\ \Eprint
  {http://arxiv.org/abs/astro-ph/0306053} {astro-ph/0306053} \BibitemShut
  {NoStop}%
\bibitem [{\citenamefont {{Shiraishi}}\ \emph {et~al.}(2017)\citenamefont
  {{Shiraishi}}, \citenamefont {{Sugiyama}},\ and\ \citenamefont
  {{Okumura}}}]{Shiraishietal:17}%
  \BibitemOpen
  \bibfield  {author} {\bibinfo {author} {\bibfnamefont {M.}~\bibnamefont
  {{Shiraishi}}}, \bibinfo {author} {\bibfnamefont {N.~S.}\ \bibnamefont
  {{Sugiyama}}}, \ and\ \bibinfo {author} {\bibfnamefont {T.}~\bibnamefont
  {{Okumura}}},\ }\href {\doibase 10.1103/PhysRevD.95.063508} {\bibfield
  {journal} {\bibinfo  {journal} {\prd}\ }\textbf {\bibinfo {volume} {95}},\
  \bibinfo {eid} {063508} (\bibinfo {year} {2017})},\ \Eprint
  {http://arxiv.org/abs/1612.02645} {arXiv:1612.02645} \BibitemShut {NoStop}%
\bibitem [{\citenamefont {{Goroff}}\ \emph {et~al.}(1986)\citenamefont
  {{Goroff}}, \citenamefont {{Grinstein}}, \citenamefont {{Rey}},\ and\
  \citenamefont {{Wise}}}]{Goroffetal:86}%
  \BibitemOpen
  \bibfield  {author} {\bibinfo {author} {\bibfnamefont {M.~H.}\ \bibnamefont
  {{Goroff}}}, \bibinfo {author} {\bibfnamefont {B.}~\bibnamefont
  {{Grinstein}}}, \bibinfo {author} {\bibfnamefont {S.-J.}\ \bibnamefont
  {{Rey}}}, \ and\ \bibinfo {author} {\bibfnamefont {M.~B.}\ \bibnamefont
  {{Wise}}},\ }\href {\doibase 10.1086/164749} {\bibfield  {journal} {\bibinfo
  {journal} {\apj}\ }\textbf {\bibinfo {volume} {311}},\ \bibinfo {pages} {6}
  (\bibinfo {year} {1986})}\BibitemShut {NoStop}%
\bibitem [{\citenamefont {{Makino}}\ \emph {et~al.}(1992)\citenamefont
  {{Makino}}, \citenamefont {{Sasaki}},\ and\ \citenamefont
  {{Suto}}}]{Makinoetal:92}%
  \BibitemOpen
  \bibfield  {author} {\bibinfo {author} {\bibfnamefont {N.}~\bibnamefont
  {{Makino}}}, \bibinfo {author} {\bibfnamefont {M.}~\bibnamefont {{Sasaki}}},
  \ and\ \bibinfo {author} {\bibfnamefont {Y.}~\bibnamefont {{Suto}}},\ }\href
  {\doibase 10.1103/PhysRevD.46.585} {\bibfield  {journal} {\bibinfo  {journal}
  {\prd}\ }\textbf {\bibinfo {volume} {46}},\ \bibinfo {pages} {585} (\bibinfo
  {year} {1992})}\BibitemShut {NoStop}%
\bibitem [{\citenamefont {{Jain}}\ and\ \citenamefont
  {{Bertschinger}}(1994)}]{JainBertschinger:94}%
  \BibitemOpen
  \bibfield  {author} {\bibinfo {author} {\bibfnamefont {B.}~\bibnamefont
  {{Jain}}}\ and\ \bibinfo {author} {\bibfnamefont {E.}~\bibnamefont
  {{Bertschinger}}},\ }\href {\doibase 10.1086/174502} {\bibfield  {journal}
  {\bibinfo  {journal} {\apj}\ }\textbf {\bibinfo {volume} {431}},\ \bibinfo
  {pages} {495} (\bibinfo {year} {1994})},\ \Eprint
  {http://arxiv.org/abs/astro-ph/9311070} {astro-ph/9311070} \BibitemShut
  {NoStop}%
\bibitem [{\citenamefont {{Hivon}}\ \emph {et~al.}(1995)\citenamefont
  {{Hivon}}, \citenamefont {{Bouchet}}, \citenamefont {{Colombi}},\ and\
  \citenamefont {{Juszkiewicz}}}]{Hivonetal:95}%
  \BibitemOpen
  \bibfield  {author} {\bibinfo {author} {\bibfnamefont {E.}~\bibnamefont
  {{Hivon}}}, \bibinfo {author} {\bibfnamefont {F.~R.}\ \bibnamefont
  {{Bouchet}}}, \bibinfo {author} {\bibfnamefont {S.}~\bibnamefont
  {{Colombi}}}, \ and\ \bibinfo {author} {\bibfnamefont {R.}~\bibnamefont
  {{Juszkiewicz}}},\ }\href@noop {} {\bibfield  {journal} {\bibinfo  {journal}
  {\aap}\ }\textbf {\bibinfo {volume} {298}},\ \bibinfo {pages} {643} (\bibinfo
  {year} {1995})},\ \Eprint {http://arxiv.org/abs/astro-ph/9407049}
  {astro-ph/9407049} \BibitemShut {NoStop}%
\bibitem [{\citenamefont {{Verde}}\ \emph {et~al.}(1998)\citenamefont
  {{Verde}}, \citenamefont {{Heavens}}, \citenamefont {{Matarrese}},\ and\
  \citenamefont {{Moscardini}}}]{Verdeetal:98}%
  \BibitemOpen
  \bibfield  {author} {\bibinfo {author} {\bibfnamefont {L.}~\bibnamefont
  {{Verde}}}, \bibinfo {author} {\bibfnamefont {A.~F.}\ \bibnamefont
  {{Heavens}}}, \bibinfo {author} {\bibfnamefont {S.}~\bibnamefont
  {{Matarrese}}}, \ and\ \bibinfo {author} {\bibfnamefont {L.}~\bibnamefont
  {{Moscardini}}},\ }\href {\doibase 10.1046/j.1365-8711.1998.01937.x}
  {\bibfield  {journal} {\bibinfo  {journal} {\mnras}\ }\textbf {\bibinfo
  {volume} {300}},\ \bibinfo {pages} {747} (\bibinfo {year} {1998})},\ \Eprint
  {http://arxiv.org/abs/astro-ph/9806028} {astro-ph/9806028} \BibitemShut
  {NoStop}%
\bibitem [{\citenamefont {{Scoccimarro}}\ \emph {et~al.}(1999)\citenamefont
  {{Scoccimarro}}, \citenamefont {{Couchman}},\ and\ \citenamefont
  {{Frieman}}}]{Scoccimarroetal:99}%
  \BibitemOpen
  \bibfield  {author} {\bibinfo {author} {\bibfnamefont {R.}~\bibnamefont
  {{Scoccimarro}}}, \bibinfo {author} {\bibfnamefont {H.~M.~P.}\ \bibnamefont
  {{Couchman}}}, \ and\ \bibinfo {author} {\bibfnamefont {J.~A.}\ \bibnamefont
  {{Frieman}}},\ }\href {\doibase 10.1086/307220} {\bibfield  {journal}
  {\bibinfo  {journal} {\apj}\ }\textbf {\bibinfo {volume} {517}},\ \bibinfo
  {pages} {531} (\bibinfo {year} {1999})},\ \Eprint
  {http://arxiv.org/abs/astro-ph/9808305} {astro-ph/9808305} \BibitemShut
  {NoStop}%
\bibitem [{\citenamefont {{Scoccimarro}}(2004)}]{Scoccimarro:04}%
  \BibitemOpen
  \bibfield  {author} {\bibinfo {author} {\bibfnamefont {R.}~\bibnamefont
  {{Scoccimarro}}},\ }\href {\doibase 10.1103/PhysRevD.70.083007} {\bibfield
  {journal} {\bibinfo  {journal} {\prd}\ }\textbf {\bibinfo {volume} {70}},\
  \bibinfo {pages} {083007} (\bibinfo {year} {2004})},\ \Eprint
  {http://arxiv.org/abs/arXiv:astro-ph/0407214} {arXiv:astro-ph/0407214}
  \BibitemShut {NoStop}%
\bibitem [{\citenamefont {{McDonald}}\ and\ \citenamefont
  {{Roy}}(2009)}]{McDonaldRoy:09}%
  \BibitemOpen
  \bibfield  {author} {\bibinfo {author} {\bibfnamefont {P.}~\bibnamefont
  {{McDonald}}}\ and\ \bibinfo {author} {\bibfnamefont {A.}~\bibnamefont
  {{Roy}}},\ }\href {\doibase 10.1088/1475-7516/2009/08/020} {\bibfield
  {journal} {\bibinfo  {journal} {\jcap}\ }\textbf {\bibinfo {volume} {8}},\
  \bibinfo {eid} {020} (\bibinfo {year} {2009})},\ \Eprint
  {http://arxiv.org/abs/0902.0991} {arXiv:0902.0991 [astro-ph.CO]} \BibitemShut
  {NoStop}%
\bibitem [{\citenamefont {{Chan}}\ \emph {et~al.}(2012)\citenamefont {{Chan}},
  \citenamefont {{Scoccimarro}},\ and\ \citenamefont {{Sheth}}}]{Chanetal:12}%
  \BibitemOpen
  \bibfield  {author} {\bibinfo {author} {\bibfnamefont {K.~C.}\ \bibnamefont
  {{Chan}}}, \bibinfo {author} {\bibfnamefont {R.}~\bibnamefont
  {{Scoccimarro}}}, \ and\ \bibinfo {author} {\bibfnamefont {R.~K.}\
  \bibnamefont {{Sheth}}},\ }\href {\doibase 10.1103/PhysRevD.85.083509}
  {\bibfield  {journal} {\bibinfo  {journal} {\prd}\ }\textbf {\bibinfo
  {volume} {85}},\ \bibinfo {eid} {083509} (\bibinfo {year} {2012})},\ \Eprint
  {http://arxiv.org/abs/1201.3614} {arXiv:1201.3614 [astro-ph.CO]} \BibitemShut
  {NoStop}%
\bibitem [{\citenamefont {{Saito}}\ \emph {et~al.}(2014)\citenamefont
  {{Saito}}, \citenamefont {{Baldauf}}, \citenamefont {{Vlah}}, \citenamefont
  {{Seljak}}, \citenamefont {{Okumura}},\ and\ \citenamefont
  {{McDonald}}}]{Saitoetal:14}%
  \BibitemOpen
  \bibfield  {author} {\bibinfo {author} {\bibfnamefont {S.}~\bibnamefont
  {{Saito}}}, \bibinfo {author} {\bibfnamefont {T.}~\bibnamefont {{Baldauf}}},
  \bibinfo {author} {\bibfnamefont {Z.}~\bibnamefont {{Vlah}}}, \bibinfo
  {author} {\bibfnamefont {U.}~\bibnamefont {{Seljak}}}, \bibinfo {author}
  {\bibfnamefont {T.}~\bibnamefont {{Okumura}}}, \ and\ \bibinfo {author}
  {\bibfnamefont {P.}~\bibnamefont {{McDonald}}},\ }\href {\doibase
  10.1103/PhysRevD.90.123522} {\bibfield  {journal} {\bibinfo  {journal}
  {\prd}\ }\textbf {\bibinfo {volume} {90}},\ \bibinfo {eid} {123522} (\bibinfo
  {year} {2014})},\ \Eprint {http://arxiv.org/abs/1405.1447} {arXiv:1405.1447}
  \BibitemShut {NoStop}%
\bibitem [{\citenamefont {{Chiang}}\ \emph {et~al.}(2014)\citenamefont
  {{Chiang}}, \citenamefont {{Wagner}}, \citenamefont {{Schmidt}},\ and\
  \citenamefont {{Komatsu}}}]{Chiangetal:14}%
  \BibitemOpen
  \bibfield  {author} {\bibinfo {author} {\bibfnamefont {C.-T.}\ \bibnamefont
  {{Chiang}}}, \bibinfo {author} {\bibfnamefont {C.}~\bibnamefont {{Wagner}}},
  \bibinfo {author} {\bibfnamefont {F.}~\bibnamefont {{Schmidt}}}, \ and\
  \bibinfo {author} {\bibfnamefont {E.}~\bibnamefont {{Komatsu}}},\ }\href@noop
  {} {\bibfield  {journal} {\bibinfo  {journal} {ArXiv e-prints}\ } (\bibinfo
  {year} {2014})},\ \Eprint {http://arxiv.org/abs/1403.3411} {arXiv:1403.3411
  [astro-ph.CO]} \BibitemShut {NoStop}%
\bibitem [{\citenamefont {{Nishimichi}}\ and\ \citenamefont
  {{Valageas}}(2015)}]{NishimichiValageas:15}%
  \BibitemOpen
  \bibfield  {author} {\bibinfo {author} {\bibfnamefont {T.}~\bibnamefont
  {{Nishimichi}}}\ and\ \bibinfo {author} {\bibfnamefont {P.}~\bibnamefont
  {{Valageas}}},\ }\href {\doibase 10.1103/PhysRevD.92.123510} {\bibfield
  {journal} {\bibinfo  {journal} {\prd}\ }\textbf {\bibinfo {volume} {92}},\
  \bibinfo {eid} {123510} (\bibinfo {year} {2015})},\ \Eprint
  {http://arxiv.org/abs/1503.06036} {arXiv:1503.06036} \BibitemShut {NoStop}%
\bibitem [{\citenamefont {{Seo}}\ and\ \citenamefont
  {{Eisenstein}}(2007)}]{SeoEisenstein:07}%
  \BibitemOpen
  \bibfield  {author} {\bibinfo {author} {\bibfnamefont {H.-J.}\ \bibnamefont
  {{Seo}}}\ and\ \bibinfo {author} {\bibfnamefont {D.~J.}\ \bibnamefont
  {{Eisenstein}}},\ }\href {\doibase 10.1086/519549} {\bibfield  {journal}
  {\bibinfo  {journal} {\apj}\ }\textbf {\bibinfo {volume} {665}},\ \bibinfo
  {pages} {14} (\bibinfo {year} {2007})},\ \Eprint
  {http://arxiv.org/abs/astro-ph/0701079} {astro-ph/0701079} \BibitemShut
  {NoStop}%
\bibitem [{\citenamefont {{Eisenstein}}\ \emph {et~al.}(2007)\citenamefont
  {{Eisenstein}}, \citenamefont {{Seo}}, \citenamefont {{Sirko}},\ and\
  \citenamefont {{Spergel}}}]{Eisensteinetal:07}%
  \BibitemOpen
  \bibfield  {author} {\bibinfo {author} {\bibfnamefont {D.~J.}\ \bibnamefont
  {{Eisenstein}}}, \bibinfo {author} {\bibfnamefont {H.-J.}\ \bibnamefont
  {{Seo}}}, \bibinfo {author} {\bibfnamefont {E.}~\bibnamefont {{Sirko}}}, \
  and\ \bibinfo {author} {\bibfnamefont {D.~N.}\ \bibnamefont {{Spergel}}},\
  }\href {\doibase 10.1086/518712} {\bibfield  {journal} {\bibinfo  {journal}
  {\apj}\ }\textbf {\bibinfo {volume} {664}},\ \bibinfo {pages} {675} (\bibinfo
  {year} {2007})},\ \Eprint {http://arxiv.org/abs/arXiv:astro-ph/0604362}
  {arXiv:astro-ph/0604362} \BibitemShut {NoStop}%
\bibitem [{\citenamefont {{Komatsu}}\ \emph {et~al.}(2011)\citenamefont
  {{Komatsu}}, \citenamefont {{Smith}}, \citenamefont {{Dunkley}},
  \citenamefont {{Bennett}}, \citenamefont {{Gold}}, \citenamefont {{Hinshaw}},
  \citenamefont {{Jarosik}}, \citenamefont {{Larson}}, \citenamefont {{Nolta}},
  \citenamefont {{Page}}, \citenamefont {{Spergel}}, \citenamefont {{Halpern}},
  \citenamefont {{Hill}}, \citenamefont {{Kogut}}, \citenamefont {{Limon}},
  \citenamefont {{Meyer}}, \citenamefont {{Odegard}}, \citenamefont {{Tucker}},
  \citenamefont {{Weiland}}, \citenamefont {{Wollack}},\ and\ \citenamefont
  {{Wright}}}]{WMAP7}%
  \BibitemOpen
  \bibfield  {author} {\bibinfo {author} {\bibfnamefont {E.}~\bibnamefont
  {{Komatsu}}}, \bibinfo {author} {\bibfnamefont {K.~M.}\ \bibnamefont
  {{Smith}}}, \bibinfo {author} {\bibfnamefont {J.}~\bibnamefont {{Dunkley}}},
  \bibinfo {author} {\bibfnamefont {C.~L.}\ \bibnamefont {{Bennett}}}, \bibinfo
  {author} {\bibfnamefont {B.}~\bibnamefont {{Gold}}}, \bibinfo {author}
  {\bibfnamefont {G.}~\bibnamefont {{Hinshaw}}}, \bibinfo {author}
  {\bibfnamefont {N.}~\bibnamefont {{Jarosik}}}, \bibinfo {author}
  {\bibfnamefont {D.}~\bibnamefont {{Larson}}}, \bibinfo {author}
  {\bibfnamefont {M.~R.}\ \bibnamefont {{Nolta}}}, \bibinfo {author}
  {\bibfnamefont {L.}~\bibnamefont {{Page}}}, \bibinfo {author} {\bibfnamefont
  {D.~N.}\ \bibnamefont {{Spergel}}}, \bibinfo {author} {\bibfnamefont
  {M.}~\bibnamefont {{Halpern}}}, \bibinfo {author} {\bibfnamefont {R.~S.}\
  \bibnamefont {{Hill}}}, \bibinfo {author} {\bibfnamefont {A.}~\bibnamefont
  {{Kogut}}}, \bibinfo {author} {\bibfnamefont {M.}~\bibnamefont {{Limon}}},
  \bibinfo {author} {\bibfnamefont {S.~S.}\ \bibnamefont {{Meyer}}}, \bibinfo
  {author} {\bibfnamefont {N.}~\bibnamefont {{Odegard}}}, \bibinfo {author}
  {\bibfnamefont {G.~S.}\ \bibnamefont {{Tucker}}}, \bibinfo {author}
  {\bibfnamefont {J.~L.}\ \bibnamefont {{Weiland}}}, \bibinfo {author}
  {\bibfnamefont {E.}~\bibnamefont {{Wollack}}}, \ and\ \bibinfo {author}
  {\bibfnamefont {E.~L.}\ \bibnamefont {{Wright}}},\ }\href {\doibase
  10.1088/0067-0049/192/2/18} {\bibfield  {journal} {\bibinfo  {journal}
  {\apjs}\ }\textbf {\bibinfo {volume} {192}},\ \bibinfo {eid} {18} (\bibinfo
  {year} {2011})},\ \Eprint {http://arxiv.org/abs/1001.4538} {arXiv:1001.4538
  [astro-ph.CO]} \BibitemShut {NoStop}%
\bibitem [{\citenamefont {{Padmanabhan}}\ \emph {et~al.}(2012)\citenamefont
  {{Padmanabhan}}, \citenamefont {{Xu}}, \citenamefont {{Eisenstein}},
  \citenamefont {{Scalzo}}, \citenamefont {{Cuesta}}, \citenamefont {{Mehta}},\
  and\ \citenamefont {{Kazin}}}]{Padmanabhanetal:12}%
  \BibitemOpen
  \bibfield  {author} {\bibinfo {author} {\bibfnamefont {N.}~\bibnamefont
  {{Padmanabhan}}}, \bibinfo {author} {\bibfnamefont {X.}~\bibnamefont {{Xu}}},
  \bibinfo {author} {\bibfnamefont {D.~J.}\ \bibnamefont {{Eisenstein}}},
  \bibinfo {author} {\bibfnamefont {R.}~\bibnamefont {{Scalzo}}}, \bibinfo
  {author} {\bibfnamefont {A.~J.}\ \bibnamefont {{Cuesta}}}, \bibinfo {author}
  {\bibfnamefont {K.~T.}\ \bibnamefont {{Mehta}}}, \ and\ \bibinfo {author}
  {\bibfnamefont {E.}~\bibnamefont {{Kazin}}},\ }\href {\doibase
  10.1111/j.1365-2966.2012.21888.x} {\bibfield  {journal} {\bibinfo  {journal}
  {\mnras}\ }\textbf {\bibinfo {volume} {427}},\ \bibinfo {pages} {2132}
  (\bibinfo {year} {2012})},\ \Eprint {http://arxiv.org/abs/1202.0090}
  {arXiv:1202.0090} \BibitemShut {NoStop}%
\bibitem [{\citenamefont {{Wagner}}\ \emph {et~al.}(2015)\citenamefont
  {{Wagner}}, \citenamefont {{Schmidt}}, \citenamefont {{Chiang}},\ and\
  \citenamefont {{Komatsu}}}]{Wagneretal:15}%
  \BibitemOpen
  \bibfield  {author} {\bibinfo {author} {\bibfnamefont {C.}~\bibnamefont
  {{Wagner}}}, \bibinfo {author} {\bibfnamefont {F.}~\bibnamefont {{Schmidt}}},
  \bibinfo {author} {\bibfnamefont {C.-T.}\ \bibnamefont {{Chiang}}}, \ and\
  \bibinfo {author} {\bibfnamefont {E.}~\bibnamefont {{Komatsu}}},\ }\href
  {\doibase 10.1093/mnrasl/slu187} {\bibfield  {journal} {\bibinfo  {journal}
  {\mnras}\ }\textbf {\bibinfo {volume} {448}},\ \bibinfo {pages} {L11}
  (\bibinfo {year} {2015})},\ \Eprint {http://arxiv.org/abs/1409.6294}
  {arXiv:1409.6294} \BibitemShut {NoStop}%
\bibitem [{\citenamefont {{Baldauf}}\ \emph {et~al.}(2016)\citenamefont
  {{Baldauf}}, \citenamefont {{Seljak}}, \citenamefont {{Senatore}},\ and\
  \citenamefont {{Zaldarriaga}}}]{Baldaufetal:16}%
  \BibitemOpen
  \bibfield  {author} {\bibinfo {author} {\bibfnamefont {T.}~\bibnamefont
  {{Baldauf}}}, \bibinfo {author} {\bibfnamefont {U.}~\bibnamefont {{Seljak}}},
  \bibinfo {author} {\bibfnamefont {L.}~\bibnamefont {{Senatore}}}, \ and\
  \bibinfo {author} {\bibfnamefont {M.}~\bibnamefont {{Zaldarriaga}}},\ }\href
  {\doibase 10.1088/1475-7516/2016/09/007} {\bibfield  {journal} {\bibinfo
  {journal} {\jcap}\ }\textbf {\bibinfo {volume} {9}},\ \bibinfo {eid} {007}
  (\bibinfo {year} {2016})},\ \Eprint {http://arxiv.org/abs/1511.01465}
  {arXiv:1511.01465} \BibitemShut {NoStop}%
\bibitem [{\citenamefont {{Jeong}}\ and\ \citenamefont
  {{Kamionkowski}}(2012)}]{JeongKamionkowski:12}%
  \BibitemOpen
  \bibfield  {author} {\bibinfo {author} {\bibfnamefont {D.}~\bibnamefont
  {{Jeong}}}\ and\ \bibinfo {author} {\bibfnamefont {M.}~\bibnamefont
  {{Kamionkowski}}},\ }\href {\doibase 10.1103/PhysRevLett.108.251301}
  {\bibfield  {journal} {\bibinfo  {journal} {Physical Review Letters}\
  }\textbf {\bibinfo {volume} {108}},\ \bibinfo {eid} {251301} (\bibinfo {year}
  {2012})},\ \Eprint {http://arxiv.org/abs/1203.0302} {arXiv:1203.0302}
  \BibitemShut {NoStop}%
\bibitem [{\citenamefont {{Byrnes}}\ \emph {et~al.}(2016)\citenamefont
  {{Byrnes}}, \citenamefont {{Dom{\`e}nech}}, \citenamefont {{Sasaki}},\ and\
  \citenamefont {{Takahashi}}}]{Byrnesetal:16}%
  \BibitemOpen
  \bibfield  {author} {\bibinfo {author} {\bibfnamefont {C.}~\bibnamefont
  {{Byrnes}}}, \bibinfo {author} {\bibfnamefont {G.}~\bibnamefont
  {{Dom{\`e}nech}}}, \bibinfo {author} {\bibfnamefont {M.}~\bibnamefont
  {{Sasaki}}}, \ and\ \bibinfo {author} {\bibfnamefont {T.}~\bibnamefont
  {{Takahashi}}},\ }\href {\doibase 10.1088/1475-7516/2016/12/020} {\bibfield
  {journal} {\bibinfo  {journal} {\jcap}\ }\textbf {\bibinfo {volume} {12}},\
  \bibinfo {eid} {020} (\bibinfo {year} {2016})},\ \Eprint
  {http://arxiv.org/abs/1610.02650} {arXiv:1610.02650} \BibitemShut {NoStop}%
\bibitem [{\citenamefont {{Hu}}\ \emph {et~al.}(2016)\citenamefont {{Hu}},
  \citenamefont {{Chiang}}, \citenamefont {{Li}},\ and\ \citenamefont
  {{LoVerde}}}]{Huetal16}%
  \BibitemOpen
  \bibfield  {author} {\bibinfo {author} {\bibfnamefont {W.}~\bibnamefont
  {{Hu}}}, \bibinfo {author} {\bibfnamefont {C.-T.}\ \bibnamefont {{Chiang}}},
  \bibinfo {author} {\bibfnamefont {Y.}~\bibnamefont {{Li}}}, \ and\ \bibinfo
  {author} {\bibfnamefont {M.}~\bibnamefont {{LoVerde}}},\ }\href {\doibase
  10.1103/PhysRevD.94.023002} {\bibfield  {journal} {\bibinfo  {journal}
  {\prd}\ }\textbf {\bibinfo {volume} {94}},\ \bibinfo {eid} {023002} (\bibinfo
  {year} {2016})},\ \Eprint {http://arxiv.org/abs/1605.01412}
  {arXiv:1605.01412} \BibitemShut {NoStop}%
\bibitem [{\citenamefont {{Chisari}}\ \emph {et~al.}(2016)\citenamefont
  {{Chisari}}, \citenamefont {{Dvorkin}}, \citenamefont {{Schmidt}},\ and\
  \citenamefont {{Spergel}}}]{Chisarietal:16}%
  \BibitemOpen
  \bibfield  {author} {\bibinfo {author} {\bibfnamefont {N.~E.}\ \bibnamefont
  {{Chisari}}}, \bibinfo {author} {\bibfnamefont {C.}~\bibnamefont
  {{Dvorkin}}}, \bibinfo {author} {\bibfnamefont {F.}~\bibnamefont
  {{Schmidt}}}, \ and\ \bibinfo {author} {\bibfnamefont {D.~N.}\ \bibnamefont
  {{Spergel}}},\ }\href {\doibase 10.1103/PhysRevD.94.123507} {\bibfield
  {journal} {\bibinfo  {journal} {\prd}\ }\textbf {\bibinfo {volume} {94}},\
  \bibinfo {eid} {123507} (\bibinfo {year} {2016})},\ \Eprint
  {http://arxiv.org/abs/1607.05232} {arXiv:1607.05232} \BibitemShut {NoStop}%
\bibitem [{\citenamefont {{Okumura}}\ \emph {et~al.}(2017)\citenamefont
  {{Okumura}}, \citenamefont {{Nishimichi}}, \citenamefont {{Umetsu}},\ and\
  \citenamefont {{Osato}}}]{Okumuraetal:17}%
  \BibitemOpen
  \bibfield  {author} {\bibinfo {author} {\bibfnamefont {T.}~\bibnamefont
  {{Okumura}}}, \bibinfo {author} {\bibfnamefont {T.}~\bibnamefont
  {{Nishimichi}}}, \bibinfo {author} {\bibfnamefont {K.}~\bibnamefont
  {{Umetsu}}}, \ and\ \bibinfo {author} {\bibfnamefont {K.}~\bibnamefont
  {{Osato}}},\ }\href@noop {} {\bibfield  {journal} {\bibinfo  {journal} {ArXiv
  e-prints}\ } (\bibinfo {year} {2017})},\ \Eprint
  {http://arxiv.org/abs/1706.08860} {arXiv:1706.08860} \BibitemShut {NoStop}%
\end{thebibliography}%

\end{document}